\documentclass[pra,showpacs,showkeys]{revtex4}% Physical Review A

\baselineskip 
\usepackage{amsmath}
\usepackage{amsfonts}            
\usepackage{amssymb}

\usepackage{mathrsfs}

\usepackage{eufrak}

\usepackage{verbatim} % for big comments

\usepackage{graphicx}% Include figure files
\usepackage{epstopdf}% Include figure files

\usepackage{dcolumn}% Align table columns on decimal point
\usepackage{bm}% bold math

\begin{document}

\title{Collisions of paramagnetic molecules in magnetic fields:  an analytic model based on Fraunhofer diffraction of matter waves}

\author{Mikhail Lemeshko}
%\email{lemeshko@fhi-berlin.mpg.de}

\author{Bretislav Friedrich}
%\email{brich@fhi-berlin.mpg.de}

\affiliation{%
Fritz-Haber-Institut der Max-Planck-Gesellschaft, Faradayweg 4-6, D-14195 Berlin, Germany
}%

\date{\today}% It is always \today, today,

\begin{abstract}

We investigate the effects of a magnetic field on the dynamics of rotationally inelastic collisions of open-shell molecules ($^2\Sigma$, $^3\Sigma$, and $^2\Pi$) with closed-shell atoms. Our treatment makes use of the Fraunhofer model of matter wave scattering and its recent extension to collisions in electric [M. Lemeshko and B. Friedrich, J.~Chem.~Phys.~\textbf{129}, 024301 (2008)] and radiative fields [M. Lemeshko and B. Friedrich, Int.~J.~Mass.~Spec.~in press (2008)]. A magnetic field aligns the molecule in the space-fixed frame and thereby alters the effective shape of the diffraction target. This significantly affects the differential and integral scattering cross sections. We exemplify our treatment by evaluating the magnetic-field-dependent scattering characteristics of the He -- CaH ($X ^2\Sigma^+$), He -- O$_2$ ($X ^3\Sigma^-$) and He -- OH ($X ^2\Pi_{\Omega}$) systems at thermal collision energies. 
Since the cross sections can be obtained for different orientations of the magnetic field with respect to  the relative velocity vector, the model also offers predictions about the frontal-versus-lateral steric asymmetry of the collisions. The steric asymmetry is found to be almost negligible for the He -- OH system, weak for the He -- CaH collisions, and strong for the He -- O$_2$. While odd $\Delta M$ transitions dominate the He -- OH $(J=3/2,f \to J', e/f)$ integral cross sections in a magnetic field parallel to the relative velocity vector, even $\Delta M$ transitions prevail in the case of the He -- CaH ($X ^2\Sigma^+$) and He -- O$_2$ ($X ^3\Sigma^-$) collision systems. For the latter system, the magnetic field opens inelastic channels that are closed in the absence of the field. These involve the  transitions $N=1, J=0 \to N', J' $ with $J'=N'$.

\end{abstract}

%Use showkeys class option if keyword
\maketitle

%------------------------- I N T R O ------------------------------------------
\section{Introduction}

All terrestrial processes, including collisions, take place in magnetic fields. And yet, quantitative studies of the effects that magnetic fields may exert on collision dynamics are mostly of a recent date, having been prompted by the newfashioned techniques to magnetically manipulate, control and confine paramagnetic atoms and molecules. 
Theoretical accounts of molecular collisions in magnetic fields are usually based on rigorous close-coupling treatments~\cite{Krems04}. Analytic models of such collisions are scarce, and limited to the Wigner regime, see, e.g., ref.~\cite{sadeghpour2000}. Here we present an analytic model of state-to-state rotationally inelastic collisions of closed-shell atoms with open-shell molecules in magnetic fields. The model, applicable to collisions at thermal and hyperthermal collision energies, is based on the Fraunhofer scattering of matter waves~\cite{Drozdov}--\cite{Faubel} and its recent extension to include collisions in electrostatic~\cite{LemFri1} and radiative~\cite{LemFri2} fields. The magnetic field affects the collision dynamics by aligning the molecular axis with respect to the relative velocity vector, thereby changing the effective shape of the diffraction
target. We consider open-shell molecules in the $^2\Sigma$, $^3\Sigma$, and $^2\Pi$ electronic states, whose body-fixed magnetic dipole moments are on the order of a Bohr magneton ~\cite{FriHer2sig}--\cite{WeinsteinCaH}. These states coincide with the most frequently occurring ground states of linear radicals, which are exemplified in our study by the CaH($X ^2\Sigma^+$), O$_2(X ^3\Sigma^-$), and OH($X ^2\Pi_{\Omega}$) species. We take, as the closed-shell collision partner, a He atom. Helium is a favorite buffer gas, used to thermalize molecules and radicals produced by laser ablation and other entrainment techniques \cite{Doyle}. 

The paper is organized as follows: in Section~\ref{sec:FraunApprox}, we briefly describe the field-free Fraunhofer model of matter-wave scattering. In Sections~\ref{sec:2Sigma},~\ref{sec:3Sigma},  and~\ref{sec:2Pi}, we extend the Fraunhofer model to account for scattering of open-shell molecules with closed-shell atoms in magnetic fields: in Section~\ref{sec:2Sigma}, we work out closed-form expressions for the partial and total differential and integral cross sections and the steric asymmetry of collisions between closed-shell atoms and paramagnetic $^2\Sigma$ molecules, and apply them to the $\text{He--CaH}(X ^2\Sigma, J=1/2 \to J')$ collision system; in Sec.~\ref{sec:3Sigma} we present the analytic theory for $^3\Sigma$ molecules and apply it to the $\text{He--O}_2(X ^3\Sigma, N=0, J=1 \to N', J')$ scattering; in Section~\ref{sec:2Pi} we develop the theory for collisions of $^2\Pi$ molecules and exemplify the results by treating the $\text{He--OH}(X ^2\Pi, J=3/2,f \to J',e/f)$ inelastic scattering. Finally, in Section~\ref{sec:conclusions}, we compare the results obtained for the collisions of the different molecules with helium and  draw conclusions from our study.

%------------------------- F I E L D - F R E E ------------------------------------------
\section{The Fraunhofer model of field-free scattering}
\label{sec:FraunApprox}

The Fraunhofer model of matter-wave scattering was recently described in Refs.~\cite{LemFri1} and~\cite{LemFri2}. Here we briefly summarize its main features. 

The model is based on two approximations. The first one replaces the amplitude
\begin{equation}
	\label{InelAmplSudden}
	f_{\mathfrak{i} \to \mathfrak{f}} (\vartheta) = \langle   \mathfrak{f} \vert f (\vartheta) \vert \mathfrak{i} \rangle
\end{equation}
for scattering into an angle $\vartheta$ from an initial, $\vert\mathfrak{i} \rangle$, to a final,  $\vert \mathfrak{f} \rangle$, state by the elastic scattering amplitude, $ f (\vartheta)$. This is tantamount to the energy sudden approximation, which is valid when the collision time is much smaller than the rotational period, as dictated by the inequality $\xi \ll 1$, where
\begin{equation}
	\label{MasseyParameter}
	\xi = \frac{\Delta E_\text{rot}k R_0}{2 E_\text{coll}}  \approx \frac{B k R_0}{E_\text{coll}},
\end{equation}
is the Massey parameter, see e.g. Refs.~\cite{Nikitin96},\cite{NikitinGasesBook}. Here $\Delta E_\text{rot}$ is the rotational level spacing, $B$ the rotational constant, $E_\text{coll}$ the collision energy, $k \equiv (2m E_\text{coll})^{1/2}/\hbar$ the wavenumber, $m$ the reduced mass of the collision system, and $R_0$ the radius of the scatterer.

\begin{figure*}[htbp]
\centering\includegraphics[width=8cm]{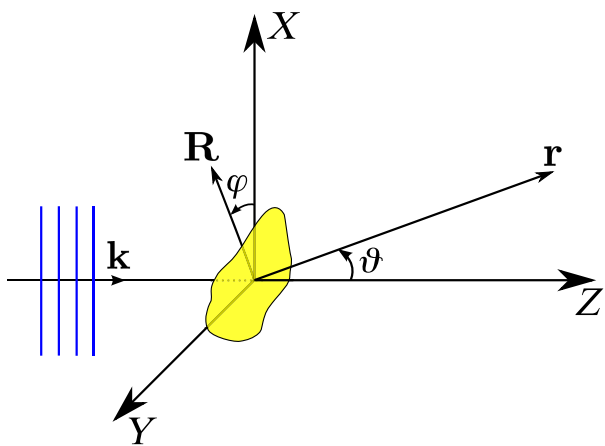}
\caption{Schematic of Fraunhofer diffraction by an impenetrable, sharp-edged obstacle as observed at a point of radius vector $\textbf{r}(X,Z)$ from the obstacle. Relevant is the shape of the obstacle in the $XY$ plane, perpendicular to the initial wave vector, $\mathbf{k}$, itself directed along the $Z$-axis of the space-fixed system $XYZ$. The angle $\varphi$ is the azimuthal angle of the radius vector $\textbf{R}$ which traces the shape of the obstacle in the $X,Y$ plane and $\vartheta$ is the scattering angle. See text.}\label{fig:fraunhofer}
\end{figure*}

The second approximation replaces the elastic scattering amplitude $ f (\vartheta)$ in Eq.~(\ref{InelAmplSudden}) by the amplitude for Fraunhofer diffraction by a sharp-edged, impenetrable obstacle as observed at a point of radiusvector $\textbf{r}$ from the scatterer, see Fig.~\ref{fig:fraunhofer}. This amplitude is given by the integral
\begin{equation}
	\label{FraunAmpl}
	f {(\vartheta)} \approx \int e^{-i k R \vartheta \cos \varphi} d \mathbf{R}
\end{equation}
Here $\varphi$ is the asimuthal angle of the radius vector $\textbf{R}$ which traces the shape of the scatterer, $R\equiv|\mathbf R|$, and $k\equiv|\mathbf k|$ with $\mathbf k$ the initial wave vector. Relevant is the shape of the obstacle in the space-fixed $XY$ plane, perpendicular to $\mathbf{k}$, itself directed along the space-fixed $Z$-axis, cf. Fig.~\ref{fig:fraunhofer}. 

We note that the notion of a sharp-edged scatterer comes close to the rigid-shell approximation, widely used in classical~\cite{Beck79}--\cite{Marks_ellips}, quantum~\cite{Bosanac}, and quasi-quantum~\cite{Stolte} treatments of field-free molecular collisions, where the collision energy by far exceeds the depth of any potential energy well.

In optics, Fraunhofer (i.e., far-field) diffraction~\cite{BornWolf} occurs when the Fresnel number is small,
\begin{equation}
	\label{FresnelNumber}
	\mathscr{F} \equiv \frac{a^2}{r \lambda} \ll 1
\end{equation}
Here $a$ is the dimension of the obstacle, $r\equiv|\textbf{r}|$ is the distance from the obstacle to the observer, and $\lambda$ is the wavelength, cf. Fig.~\ref{fig:fraunhofer}. Condition~(\ref{FresnelNumber}) is well satisfied for nuclear scattering at MeV collision energies as well as for molecular collisions at thermal and hyperthermal energies. In the latter case, inequality~(\ref{FresnelNumber}) is fulfilled due to the compensation of the larger molecular size $a$ by a larger de~Broglie wavelength $\lambda$ pertaining to thermal molecular velocities.

For a nearly-circular scatterer, with a boundary $R (\varphi) = R_0 +\delta(\varphi)$ in the $XY$ plane, the Fraunhofer integral of Eq.~(\ref{FraunAmpl}) can be evaluated and expanded in a power series in the deformation $\delta(\varphi)$,
\begin{equation}
	\label{AmplitudeExpansion}
	f_{}  {(\vartheta)}  = f_0 (\vartheta) + f_1 (\vartheta,\delta) + f_2(\vartheta,\delta^2)+\cdots
\end{equation}
with $f_0(\vartheta)$ the amplitude for scattering by a disk of radius $R_0$
\begin{equation}
	\label{AmplSphere}
	f_0 (\vartheta) = i (k R_0^2) \frac{J_1 (k R_0 \vartheta)}{(k R_0 \vartheta)}
\end{equation}
and $f_1$ the lowest-order anisotropic amplitude,
\begin{equation}
	\label{AmplFirstOrder}
	f_1(\vartheta) = \frac{i k}{2 \pi} \int_{0}^{2 \pi} \delta(\varphi) e^{- i (k R_0 \vartheta) \cos \varphi} d\varphi
\end{equation}
where $J_1$ is a Bessel function of the first kind. Both Eqs.~(\ref{AmplSphere}) and~(\ref {AmplFirstOrder}) are applicable at small values of $\vartheta \lesssim 30^{\circ}$, i.e., within the validity of the approximation $\sin \vartheta \approx \vartheta$. 

The scatterer's shape in the space fixed frame, see Fig.~\ref{fig:fraunhofer}, is given by
\begin{equation}
	\label{RhoExpSpaceFixed}
	R (\alpha, \beta, \gamma ; \theta, \varphi) = \sum_{\kappa \nu \rho} \Xi_{\kappa \nu} \mathscr{D}_{\rho \nu}^{\kappa} (\alpha \beta \gamma) Y_{\kappa \rho} (\theta, \varphi)
\end{equation}
where $(\alpha,\beta,\gamma)$ are the Euler angles through which the body-fixed frame is rotated relative to the space-fixed frame, $(\theta, \varphi)$ are the polar and azimuthal angles in the space-fixed frame, $\mathscr{D}_{\rho \nu}^{\kappa} (\alpha \beta \gamma)$ are the Wigner rotation matrices, and $\Xi_{\kappa \nu}$ are the Legendre moments describing the scatterer's shape in the body-fixed frame. Clearly, the term with  $\kappa=0$ corresponds to a disk of radius $R_0$,
\begin{equation}
	\label{R0viab}
	R_0 \approx \frac{\Xi_{00}}{\sqrt{4\pi}}
\end{equation}
Since of relevance is the shape of the target in the $XY$ plane, we set $\theta=\frac{\pi}{2}$ in Eq.~(\ref{RhoExpSpaceFixed}). As a result,
\begin{equation}
\label{deltaphi}
	\delta(\varphi)=R (\alpha, \beta, \gamma ; \tfrac{\pi}{2}, \varphi)-R_0=R (\varphi) - R_0=\underset{\kappa \neq 0 }{ \sum_{\kappa \nu \rho}} \Xi_{\kappa \nu} \mathscr{D}_{\rho \nu}^{\kappa} (\alpha \beta \gamma) Y_{\kappa \rho} (\tfrac{\pi}{2}, \varphi)
\end{equation}
By combining Eqs.~(\ref {InelAmplSudden}), (\ref{AmplFirstOrder}), and (\ref{deltaphi}) we finally obtain
\begin{equation}
	\label{InelAmplExpress}
	f_{\mathfrak{i} \to \mathfrak{f}} (\vartheta) \approx \langle \mathfrak{f} \vert f_0 + f_1 \vert \mathfrak{i} \rangle = \langle \mathfrak{f} \vert f_1 \vert \mathfrak{i} \rangle =  \frac{i k R_0}{2 \pi}  \underset{\kappa+\rho~\textrm{even}}{\underset{\kappa \neq 0 } {\sum_{\kappa \nu \rho}}} \Xi_{\kappa \nu} \langle \mathfrak{f} \vert \mathscr{D}_{\rho \nu}^{\kappa} \vert \mathfrak{i} \rangle F_{\kappa \rho} J_{\vert \rho \vert} (k R_0 \vartheta)
\end{equation}
where
\begin{equation}
	\label{Flamnu}
	F_{\kappa \rho} = \left \{ 	\begin{array}{ccl}
		(-1)^{\rho} 2\pi \left( \frac{2\kappa+1}{4\pi} \right)^{\frac{1}{2}} (-i)^{\kappa} \frac{\sqrt{(\kappa+\rho)! (\kappa-\rho)! }}{(\kappa+\rho)!! (\kappa-\rho)!! }  &    &    \textrm{ for $\kappa+\rho$ ~even~and~ $\kappa \ge \rho$} 
		\\ \\						0     &    &     \textrm{ elsewhere} 
	\end{array}    \right .
\end{equation}
For negative values of $\rho$, the factor $(-i)^{\kappa}$ is to be replaced by $i^{\kappa}$.

%------------------------- 2 S I G M A -------------------------------------------

\section{Scattering of $^2\Sigma$ molecules by closed-shell atoms in a magnetic field}
\label{sec:2Sigma}

%----------------------------------------------

\subsection{A $^2\Sigma$ molecule in a magnetic field}
\label{sec:Zeem2Sigma}

The field-free Hamiltonian of a rigid $^2\Sigma$ molecule
\begin{equation}
	\label{2SigFFHamiltonian}
	H_0 =  B N^2 + \gamma \mathbf{N \cdot S}
\end{equation}
is represented by a $2\times 2$ matrix, diagonal in the Hund's case (b) basis, $\vert N, J,M \rangle$. Here $\mathbf N$ and $\mathbf S$ are the rotational and (electronic) spin angular momenta, $B$ is the rotational constant and $\gamma$ the spin-rotation constant. Its eigenfunctions 
\begin{equation}
	\label{WFfieldfree1}
	\Psi_{\pm} (J, M)  = \frac{1}{\sqrt{2}} \biggl [ \bigl | S, {\tfrac{1}{2}} \bigr > \bigl | J, \Omega, M  \bigr > \pm  \bigl | S, -\tfrac{1}{2}  \bigr > \bigl | J, - \Omega, M \bigr > \biggr ],
\end{equation}
are combinations of (electronic) spin functions $\vert S, M_S \rangle$ with Hund's case (a) (i.e., symmetric top) functions $\vert J, \Omega,M \rangle$ pertaining to the total angular momentum $\mathbf J=\mathbf N+\mathbf S$, whose projections on the space- and body-fixed axes are $M$ and $\Omega=\pm \frac{1}{2}$, respectively. The Hund's case~(a) wavefunctions are given by:
 \begin{equation}
	\label{RotatWF}
	\vert J, M, \Omega \rangle = \sqrt{\frac{2 J+1}{4 \pi}}  \mathscr{D}_{M \Omega}^{J \ast} (\varphi, \theta, \gamma=0)
\end{equation}

The $\Psi_{+}$ and $\Psi_{-}$ states are conventionally designated as $F_1$ and $F_2$ states, for which the rotational quantum number $N=J-\frac{1}{2}$ and $N=J+\frac{1}{2}$, respectively.
Equation~(\ref{WFfieldfree1}) can be recast in terms of $N$ instead of $J$:
\begin{equation}
	\label{WFfieldfree2}
	\vert \Psi_{\epsilon} (N, M) \rangle = \frac{1}{\sqrt{2}} \biggl [ \bigl | S, \tfrac{1}{2}  \bigr > \bigl | N+ \tfrac{\epsilon}{2}, \Omega, M  \bigr > + \epsilon \bigl | S, -\tfrac{1}{2}  \bigr > \bigl | N+ \tfrac{\epsilon}{2}, - \Omega, M \bigr > \biggr ],
\end{equation}
with $\epsilon = \pm 1$.

The eigenvalues corresponding to states $F_1$ and $F_2$ are given by
 \begin{equation}
	\label{Eplus}
	E_{+} \Bigl ( N+\tfrac{1}{2}, M ; F_1 \Bigr ) = BN(N+1) + \frac{\gamma}{2} N
\end{equation}
 \begin{equation}
	\label{Eminus}
	E_{-} \Bigl ( N-\tfrac{1}{2}, M ; F_2 \Bigr ) = BN(N+1) - \frac{\gamma}{2} (N+1),
\end{equation}
whence we see that the spin-rotation interaction splits each rotational level into a doublet separated by $\Delta E \equiv E_+-E_-=\gamma (N+\frac{1}{2})$. 
%----------------------------------------------

In a static magnetic field, $\mathscr{H}$, directed along the space-fixed $Z$ axis, the Hamiltonian acquires a magnetic dipole potential which is proportional to the projection, $S_Z$, of $\mathbf S$ on the $Z$ axis
\begin{equation}
	\label{MagnDipPot1}
	V_{\text{m}} = S_Z \omega_\text{m} B,
\end{equation}
with 
\begin{equation}
	\label{omegapar2}
	\omega_\text{m} \equiv \frac{g_S \mu_B \mathscr{H}} {2B} 
\end{equation}
a dimensionless interaction parameter involving the electron gyromagnetic ratio $g_S \simeq 2.0023$, the Bohr magneton $\mu_B$, and the rotational constant $B$.

The Zeeman eigenproperties of a $^2\Sigma$ molecule can be readily obtained in closed form, since the $V_{\text m}$ operator couples states that differ in $N$ by $0$ or $\pm2$ and, therefore, the Hamiltonian matrix, $H=H_0+V_{\text m}$, factors into $2\times2$ blocks for each $N$:
\begin{equation}
	\label{HamMatr2x2}
	H = -\omega_\text{m} B \left (
	\begin{array}{c c c}
	-\frac{M}{2N+1} + \frac{E_{-}}{\omega_\text{m} B} &  & \frac{1}{2} [1 - \frac{M^2}{(N+1/2)^2}]^{\frac{1}{2}} \\ \\
	 \frac{1}{2} [1 - \frac{M^2}{(N+1/2)^2}]^{\frac{1}{2}}  &  & \frac{M}{2N+1} + \frac{E_{+}}{\omega_\text{m} B}	 \\
	\end{array}	
	\right )
\end{equation}
As a result, the Zeeman eigenfunctions of a $^2\Sigma$ molecule are given by a linear combination of the field-free wavefunctions~(\ref{WFfieldfree2}),
\begin{equation}
	\label{Psizeem}
	\psi (\tilde{N},\tilde{J},M; \omega_\text{m})  = a(\omega_\text{m})\bigl | \Psi_{-} (N,M) \bigr >  + b(\omega_\text{m})\bigl | \Psi_{+} (N,M) \bigr >,
\end{equation}
with the hybridization coefficients $a(\omega_\text{m})$ and $b(\omega_\text{m})$ obtained by diagonalizing Hamiltonian~(\ref{HamMatr2x2}). Although $N$ and $J$ are no longer good quantum numbers in the magnetic field, they can be employed as adiabatic labels of the states: we use $\tilde{N}$ and $\tilde{J}$ to denote the angular momentum quantum numbers of the field-free state that adiabatically correlates with the given state in the field. Since the Zeeman eigenfunction comprises rotational states with either $N$ even or $N$ odd, the parity of the eigenstates remains definite even in the presence of the magnetic field; it is given by $(-1)^{\tilde{N}}$.

The degree of mixing of the Hund's case (b) states that make up a $^2\Sigma$ Zeeman eigenfunction is determined by the splitting of the rotational levels measured in terms of the rotational constant,  $\Delta E/B$: for $\omega_\text{m} \le \Delta E/B$ the mixing (hybridization) is incomplete, while it is perfect in the high-field limit, $\omega_\text{m} \gg \Delta E/B$. We note that in the high-field limit, the eignevectors can be found from matrix~(\ref{HamMatr2x2}) with $E_{\pm}/{\omega_\text{m} B} \to 0$. As an example, Table~\ref{table:ab_coefs} lists the values of the hybridization coefficients $a(\omega_\text{m})$ and $b(\omega_\text{m})$ for the $N=2, J=\frac{5}{2}, M$ states of the CaH molecule in the high-field limit, which is attained at $\omega_\text{m} \gg 0.025$.

\begin{figure*}[htbp]
\centering
\includegraphics[width=8cm]{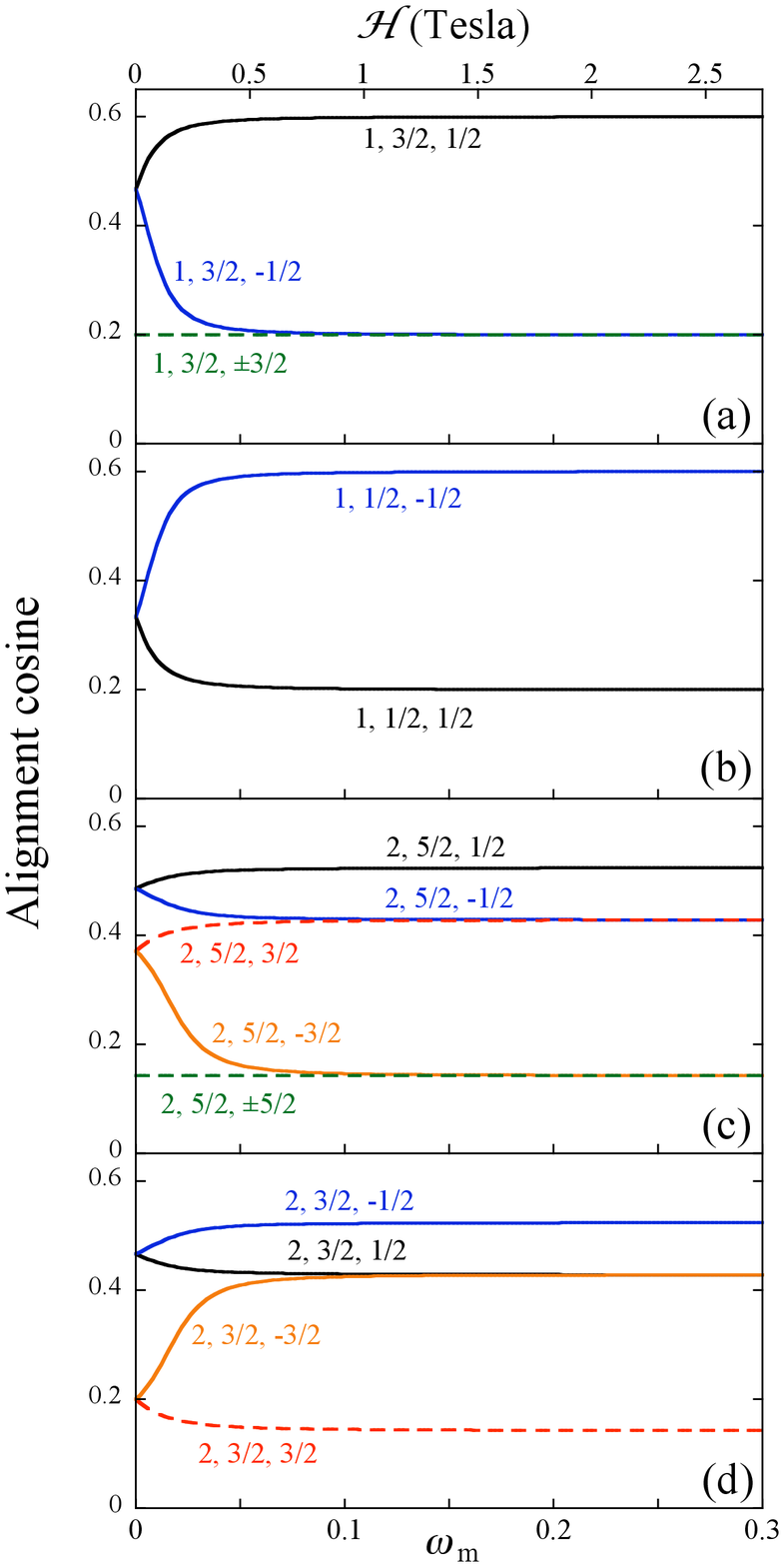}
\caption{Expectation values of the alignment cosine $\langle \cos^2\theta \rangle$ for the Zeeman states of CaH$(X^2\Sigma^+)$ as a function of the magnetic field strength parameter $\omega_\text{m}$. States are labeled as $\tilde{N},\tilde{J},M$, see text.}
\label{fig:CaH_cos2}
\end{figure*}

The degree of molecular axis alignment is given by the alignment cosine, $\langle \cos^2 \theta \rangle$, which, in the $^2\Sigma$ case, can be obtained in closed form. To the best of our knowledge, this result has not been presented in the literature before; therefore, we give it in Appendix~\ref{app:cos2}. The dependence of the alignment cosine on the magnetic field strength parameter $\omega_\text{m}$ is shown in Fig.~\ref{fig:CaH_cos2} for the two lowest $N$ states of the CaH molecule. One can see that for $\omega_\text{m} \gg \Delta E/B$, the alignment cosine smoothly approaches a constant value, corresponding to as good an alignment as the uncertainty principle allows.

%----------------------------------------------

\subsection{The field-dependent scattering amplitude}
\label{sec:Fraun2Sigma}

In what follows, we consider scattering from the $N=0, J=1/2$ state to some $N', J'$ state in a magnetic field. Since the $N=0$ state of a $^2\Sigma$ molecule is not aligned, the effects of the magnetic field on the scattering arise solely from the alignment of the final state. 

In order to account for an arbitrary direction of the electric field with respect to the initial wave vector $\mathbf{k}$, we introduce a field-fixed coordinate system $X^{\sharp} Y^{\sharp}Z^{\sharp}$, whose $Z^{\sharp}$-axis is defined by the direction of the electric field vector $\boldsymbol{\varepsilon }$. By making use of the relation
\begin{equation}
	\label{FieldToSpaceDmatrix}
	\mathscr{D}_{M \Omega}^{J \ast} (\varphi^{\sharp}, \theta^{\sharp}, 0) = \sum_{\xi} \mathscr{D}_{\xi M}^{J} (\varphi_{\varepsilon}, \theta_{\varepsilon}, 0) \mathscr{D}_{\xi \Omega}^{J \ast} (\varphi, \theta, 0) 
\end{equation}
we transform the wavefunctions~(\ref{Psizeem}) to the space-fixed frame. For the initial and the final states we have:
\begin{multline}
	\label{2SigInit}
	\vert \mathfrak{i} (N,M) \rangle = \frac{1}{\sqrt{4 \pi}} \sum_{\xi} \Biggl \{  a(\omega_\text{m}) \sqrt{N} \mathscr{D}_{\xi M}^{N-\tfrac{1}{2}} (\varphi_{\varepsilon},\theta_{\varepsilon},0)	 \biggl [ \mathscr{D}_{\xi \Omega}^{N-\tfrac{1}{2} \ast} (\varphi, \theta, 0) - \mathscr{D}_{\xi -\Omega}^{N-\tfrac{1}{2} \ast} (\varphi, \theta, 0) \biggr] \\
	+ b(\omega_\text{m}) \sqrt{N+1} \mathscr{D}_{\xi M}^{N+\tfrac{1}{2}} (\varphi_{\varepsilon},\theta_{\varepsilon},0)	 \biggl [ \mathscr{D}_{\xi \Omega}^{N+\tfrac{1}{2} \ast} (\varphi, \theta, 0) - \mathscr{D}_{\xi -\Omega}^{N+\tfrac{1}{2} \ast} (\varphi, \theta, 0) \biggr]
        \Biggr \}
\end{multline}

\begin{multline}
	\label{2SigFinal}
	\langle \mathfrak{f} (N',M') \vert = \frac{1}{\sqrt{4 \pi}} \sum_{\xi'} \Biggl \{  a'(\omega_\text{m}) \sqrt{N'} \mathscr{D}_{\xi' M'}^{N'-\tfrac{1}{2}} (\varphi_{\varepsilon},\theta_{\varepsilon},0) \biggl [ \mathscr{D}_{\xi' \Omega}^{N'-\tfrac{1}{2} \ast} (\varphi, \theta, 0) - \mathscr{D}_{\xi' -\Omega}^{N'-\tfrac{1}{2} \ast} (\varphi, \theta, 0) \biggr] \\
	+ b'(\omega_\text{m}) \sqrt{N'+1} \mathscr{D}_{\xi' M'}^{N'+\tfrac{1}{2}} (\varphi_{\varepsilon},\theta_{\varepsilon},0)	 \biggl [ \mathscr{D}_{\xi' \Omega}^{N'+\tfrac{1}{2} \ast} (\varphi, \theta, 0) - \mathscr{D}_{\xi' -\Omega}^{N'+\tfrac{1}{2} \ast} (\varphi, \theta, 0) \biggr]
        \Biggr \}
\end{multline}
where $\Omega = \frac{1}{2}$ for a $^2 \Sigma$ molecule.

By substituting from Eqs.~(\ref{2SigInit}) and~(\ref{2SigFinal}) into Eq.~(\ref{InelAmplExpress}), we finally obtain the scattering amplitude for inelastic collisions of $^2\Sigma$ molecules with closed-shell atoms in a magnetic field:
\begin{multline}
	\label{2SigScatAmplBig}	
	f_{\mathfrak{i} \to \mathfrak{f}}^{\omega_\text{m}} (\vartheta)= \frac{i k R_0}{4 \pi} \underset{\kappa+\rho~\textrm{even}}{\underset{\kappa \neq 0 } {\sum_{\kappa \rho}}} \Xi_{\kappa 0} \mathscr{D}_{-\rho, \Delta M}^{\kappa \ast}  (\varphi_{\varepsilon},\theta_{\varepsilon},0)  F_{\kappa \rho} J_{\vert \rho \vert} (k R_0 \vartheta) \Biggl[ (-1)^{\kappa} + (-1)^{\Delta N} \biggr]  \\
	\times\Biggl \{ a(\omega_\text{m}) a'(\omega_\text{m}) \sqrt{\frac{N}{N'}} C\Bigl (N-\tfrac{1}{2}, \kappa, N'-\tfrac{1}{2}; \Omega 0 \Omega \Bigr) C\Bigl (N-\tfrac{1}{2}, \kappa, N'-\tfrac{1}{2}; M \Delta M M'  \Bigr) \\
	+a(\omega_\text{m}) b'(\omega_\text{m}) \sqrt{\frac{N}{N'+1}} C\Bigl (N-\tfrac{1}{2}, \kappa, N'+\tfrac{1}{2}; \Omega 0 \Omega \Bigr) C\Bigl (N-\tfrac{1}{2}, \kappa, N'+\tfrac{1}{2}; M \Delta M M'  \Bigr) \\
	+a'(\omega_\text{m}) b(\omega_\text{m}) \sqrt{\frac{N+1}{N'}} C\Bigl (N+\tfrac{1}{2}, \kappa, N'-\tfrac{1}{2}; \Omega 0 \Omega \Bigr) C\Bigl (N+\tfrac{1}{2}, \kappa, N'-\tfrac{1}{2}; M \Delta M M'  \Bigr) \\
	+b(\omega_\text{m}) b'(\omega_\text{m}) \sqrt{\frac{N+1}{N'+1}} C\Bigl (N+\tfrac{1}{2}, \kappa, N'+\tfrac{1}{2}; \Omega 0 \Omega \Bigr) C\Bigl (N+\tfrac{1}{2}, \kappa, N'+\tfrac{1}{2}; M \Delta M M'  \Bigr)	
	\Biggr \} 
\end{multline}
As noted above, there is no hybridization of the initial state for the $N=0, J=\tfrac{1}{2} \to N', J'$ collisions, i.e., $a(\omega_\text{m})=0$, $b(\omega_\text{m})=1$ in Eq.~(\ref{2SigScatAmplBig}). By making use of the properties of the Clebsch-Gordan coefficients~\cite{Zare},\cite{Varshalovich}, the expression for the scattering amplitude from the $N=0, J=\tfrac{1}{2}, M=\pm \tfrac{1}{2}$ state to an $N', J', M'$ state simplifies to
\begin{multline}
	\label{2SigScatAmpl}	
	f_{0,\tfrac{1}{2},\pm\tfrac{1}{2} \to N',J',M'}^{\omega_\text{m}} (\vartheta)= \frac{i k R_0}{2 \pi} \frac{\Xi_{N' 0} }{2N'+1} \left \{ \underset{\rho+N' \text{even}} {\sum_{\rho}} d_{-\rho, \Delta M}^{N'} (\theta_{\varepsilon}) F_{N' \rho} J_{\vert \rho \vert} (k R_0 \vartheta)  \right \} \\
	\times  \Biggl [ \pm a'(\omega_\text{m}) \sqrt{N' \mp M' +\tfrac{1}{2} } + b'(\omega_\text{m}) \sqrt{N' \pm M' +\tfrac{1}{2} }	\Biggr ] 
\end{multline}
The amplitude is seen to be directly proportional to the $\Xi_{N' 0}$ Legendre moment. We note that the cross section for the $N,J,M \to N',J',M'$ transition differs from that for the $N,J,-M \to N',J',-M'$ scattering. This is because the magnetic field completely lifts the degeneracy of the $M$ states, in contrast to the electric field case~\cite{LemFri1}.

%------------------R E S U L T S --- F O R --- CaH ----------------------------

\subsection{Results for $\text{He -- CaH}(X ^2\Sigma, J=1/2 \to J')$ scattering in a magnetic field}
\label{He-CaH}
Here we apply the analytic model scattering to the He -- CaH($^2\Sigma^+, J=\frac{1}{2} \to J')$ collision system. The CaH molecule, employed previously in thermalization experiments with a He buffer gas~\cite{WeinsteinCaH}, \cite{FriedrichCaH}, has a rotational constant $B = 4.2766$ cm$^{-1}$ and a spin-rotational interaction parameter $\gamma = 0.0430$ cm$^{-1}$~\cite{MartinCaH}. Such values of molecular constants result in an essentially perfect mixing (and alignment) of the molecular states for field strengths $\mathscr{H}\ge 0.1$ Tesla, see Sec.~\ref{sec:Zeem2Sigma}.

\begin{figure*}[htbp]
\centering\includegraphics[width=8cm]{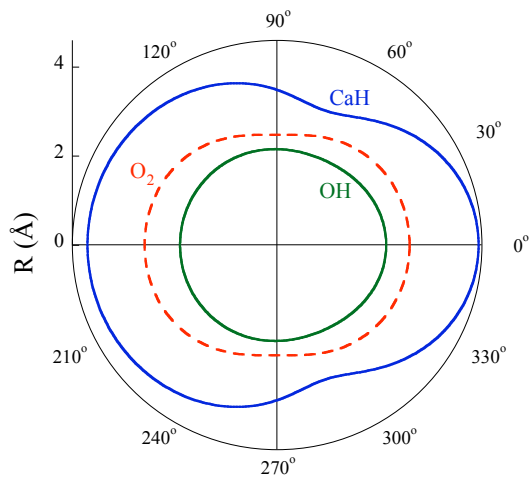}
\caption{Equipotential lines $R (\theta)$ for the He -- CaH ($X ^2 \Sigma^+$) and He -- O$_2$ ($X ^3 \Sigma^-$) potential energy surfaces at a collision energy of 200 cm$^{-1}$, and for the He -- OH ($X ^2 \Pi_{\Omega}$) potential at 1000 cm$^{-1}$. The Legendre moments for the potential energy surfaces are listed in Table \ref{table:legendre_coefs}.}\label{fig:PEScut}
\end{figure*}

 According to Ref.~\cite{PESHe-CaH}, the ground-state He--CaH potential energy surface has a global minimum of $-10.6$ cm$^{-1}$. Such a weak attractive well can be neglected at a collision energy as low as $200$ cm$^{-1}$ (which corresponds to a wave number $k=6.58$ \r{A}$^{-1}$).  The corresponding value of the Massey parameter, $\xi \approx 0.5$, warrants the validity of the sudden approximation to the He -- CaH collision system from this collision energy on. The ``hard shell" of the potential energy surface was found by a fit to Eq.~(\ref{RhoExpSpaceFixed}) for $\kappa \le 8$, and is shown in Fig.~\ref{fig:PEScut}. The coefficients $\Xi_{\kappa 0}$ obtained from the fit are listed in Table~\ref{table:legendre_coefs}. According to Eq.~(\ref{R0viab}), the $\Xi_{00}$ coefficient determines the hard-sphere radius $R_0$, which is responsible for elastic scattering.

%----------------------------------------------

\subsubsection{Differential cross sections}
\label{diffcrossCaH}

\begin{figure*}[htbp]
\centering
\includegraphics[width=8cm]{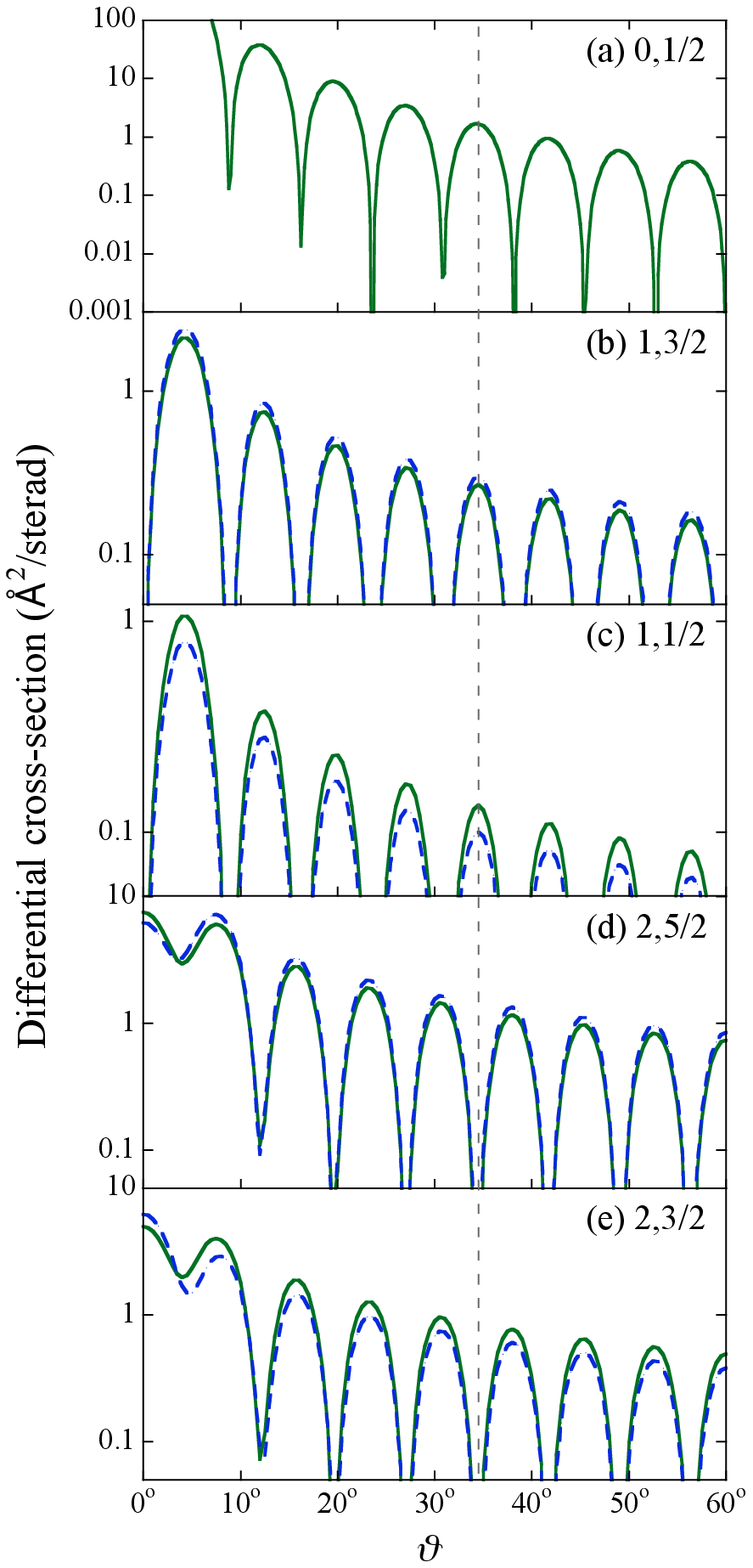}
\caption{Differential cross sections for the He -- CaH $(N=0, J=\frac{1}{2} \to N', J')$ collisions in a magnetic field $\omega_\text{m}=0.3$ (blue dashed line) parallel to the relative velocity vector, $\mathscr{H} \parallel {\bf k}$. The field-free cross sections are shown by the green solid line. The dashed vertical line serves to guide the eye in discerning the angular shifts of the partial cross sections.}
\label{fig:CaH_diff_par}
\end{figure*}

\begin{figure*}[htbp]
\includegraphics[width=8cm]{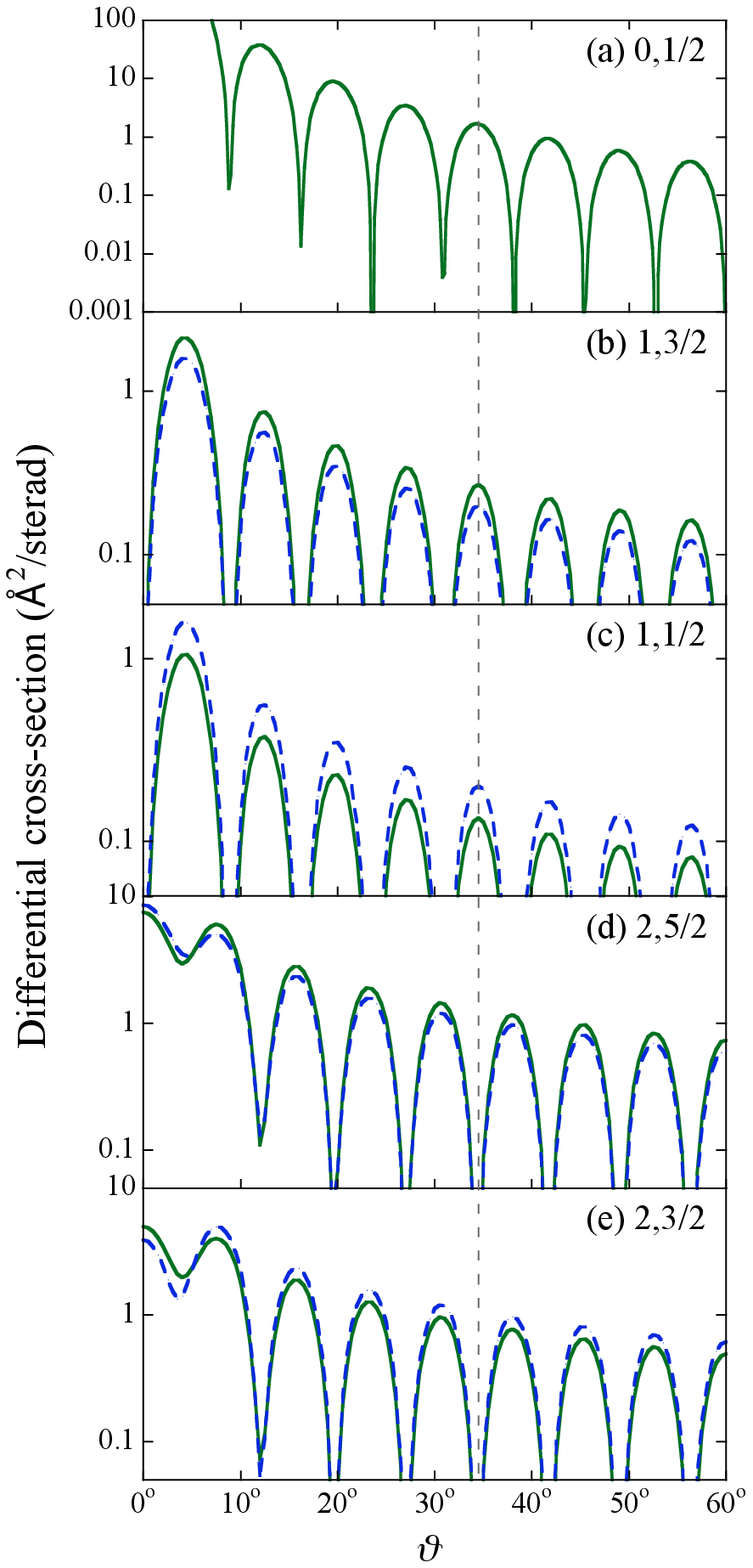}
\caption{Differential cross sections for the He -- CaH $(N=0, J=\frac{1}{2} \to N', J')$ collisions in a magnetic field $\omega_\text{m}=0.3$ (blue dashed line) perpendicular to the relative velocity vector, $\mathscr{H} \perp {\bf k}$. The field-free cross sections are shown by the green solid line. The dashed vertical line serves to guide the eye in discerning the angular shifts of the partial cross sections.}
\label{fig:CaH_diff_perp}
\end{figure*}

The state-to-state differential cross sections for scattering in a field parallel ($\parallel$) and perpendicular ($\perp$) to $\mathbf{k}$ are given by
\begin{equation}
	\label{DiffCrossFieldsJaver}
	\mathcal{I}_{0 \to J'}^{\omega_\text{m},(\parallel,\perp)}(\vartheta)=\sum_{M'} \mathcal{I}_{0,0 \to J',M'}^{\omega_\text{m},(\parallel,\perp)}(\vartheta)
\end{equation}
with
\begin{equation}
	\label{DiffCrossFieldsJM}
	\mathcal{I}_{0,0 \to J',M'}^{\omega_\text{m},(\parallel,\perp)}(\vartheta)=\left \vert f_{0,0 \to \tilde{J'}, M' }^{\omega_\text{m}, (\parallel,\perp)} (\vartheta)  \right \vert^2
\end{equation}
They are presented in Figs.~\ref{fig:CaH_diff_par},~\ref{fig:CaH_diff_perp} for He--CaH collisions at zero field, $\omega_\text{m} = 0$, as well as at high field, $\omega_\text{m} = 0.3$ (corresponding to   $\mathscr{H}$=2.75~T for CaH), where the hybridization and alignment are as complete as they can get.

From Eq.~(\ref{2SigScatAmpl}) for the scattering amplitude, we see that the differential cross section for the $N=0 \to N'$ transitions is proportional to the $\Xi_{N' 0}$ Legendre moment. According to Table~\ref{table:legendre_coefs}, the Legendre expansion of the He--CaH potential energy surface is dominated by $\Xi_{20}$. Therefore, the transition $N=0 \to N'=2$ provides the largest contribution to the cross section.

The field dependence of the scattering amplitude, Eq.~(\ref{2SigScatAmpl}), is encapsulated in the coefficients $a'(\omega_\text{m})$ and $b'(\omega_\text{m})$, whose values cannot affect the angular dependence, as this is determined solely by the Bessel functions,  $J_{\vert \rho \vert} (k R_0 \vartheta)$. Furthermore, the summation in Eq.~(\ref{2SigScatAmpl}) includes only even $\rho$ for even $N'$, and odd $\rho$ for odd $N'$. From the asymptotic properties of Bessel functions~\cite{Watson}, we have for large angles such that $\vartheta \gg \pi \rho/2 k R_0$:
\begin{equation}
	\label{AsymptBessel}
	J_{\vert \rho \vert} (k R_0 \vartheta)   \sim \left \{ 
	\begin{array}{ccl}
		\cos \left(k R_0 \vartheta - \frac{\pi}{4} \right) &    & \textrm{ for $\rho$~even } \\ \\
		\sin \left(k R_0 \vartheta - \frac{\pi}{4}  \right) &    &\textrm{ for $\rho$~odd} 
	\end{array}    \right .
\end{equation}
For the He -- CaH system, the phase shift between the $J_0$ and $J_2$ Bessel functions, which contribute to the $N=0 \to N'=1,2$ transitions, is negligibly small at angles up to about 30$^\circ$. Therefore there is no field-induced phase shift, neither in the parallel nor in the perpendicular case, as illustrated by Figs.~\ref{fig:CaH_diff_par},~\ref{fig:CaH_diff_perp}.

Figs.~\ref{fig:CaH_diff_par} and~\ref{fig:CaH_diff_perp}  show that the magnetic field induces only small changes in the amplitudes of the cross sections, without shifting their oscillations. The amplitude variation is so small because the magnetic field fails to mix contributions from the different $\Xi_{\kappa,0}$ Legendre moments, in contrast to scattering in electrostatic~\cite{LemFri1} and radiative~\cite{LemFri2} fields. The changes in the amplitudes of the differential cross sections are closely related to the field dependence of the partial integral cross sections, which are analyzed next.

%--------------------------------------------------------------------------------------------------

\subsubsection{Integral cross sections}
\label{sec:CaHIntCross}

The angular range, $\vartheta \lesssim 30^{\circ}$, where the Fraunhofer approximation applies the best, comprises the largest-impact-parameter collisions that contribute to the scattering the most, see Figs.~\ref{fig:CaH_diff_par} and~\ref{fig:CaH_diff_perp}. Therefore, the integral cross section can be obtained, to a good approximation, by integrating the Fraunhofer differential cross sections, Eq. (\ref{DiffCrossFieldsJaver}) and~(\ref{DiffCrossFieldsJM}), over the solid angle $\sin\vartheta d \vartheta d \varphi$, with  $0 \le\vartheta \le \pi$ and $0 \le \varphi \le 2\pi$. 

\begin{figure*}[htbp]
\centering
\includegraphics[width=8cm]{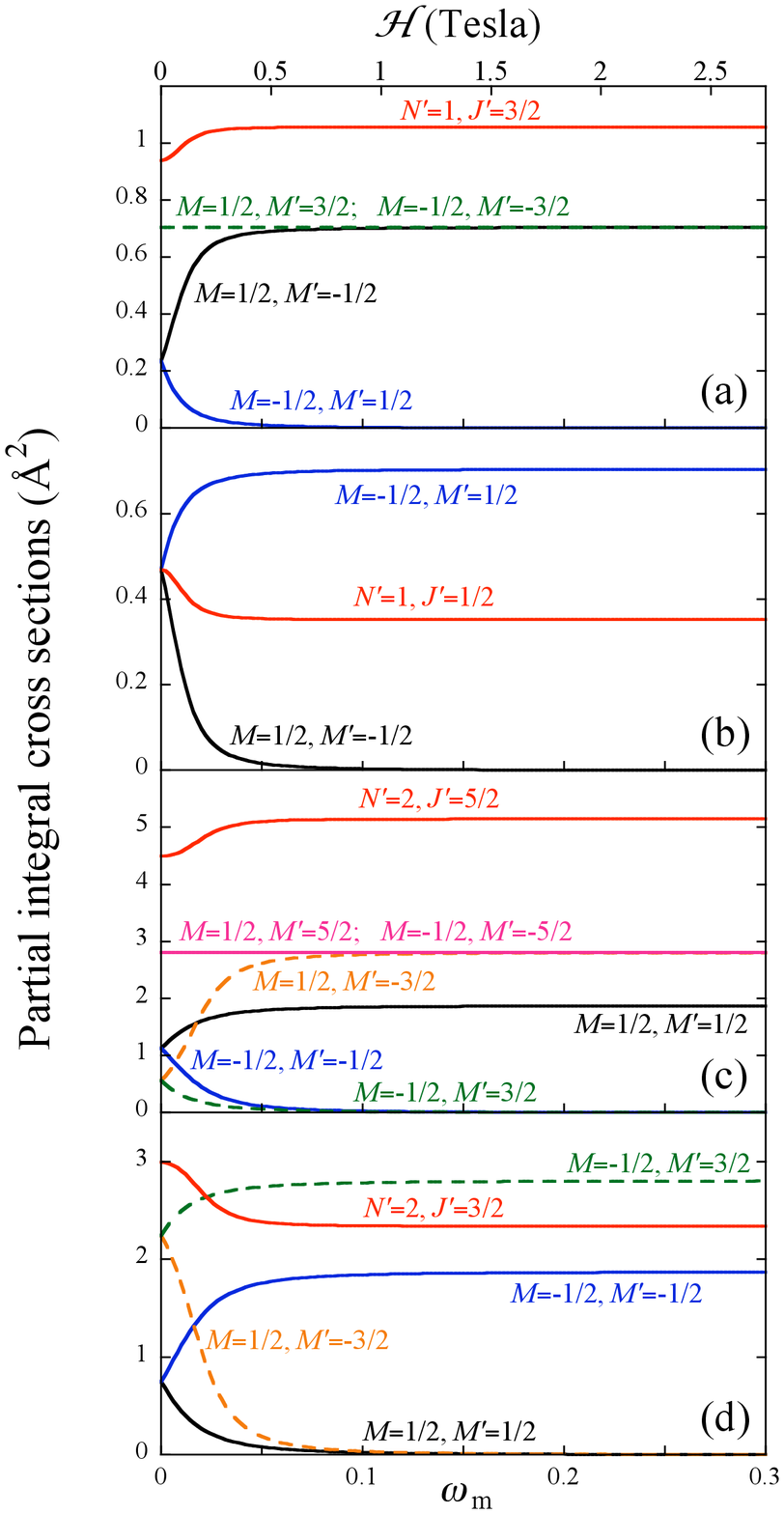}
\caption{Partial integral cross sections for the He -- CaH $(N=0, J=\frac{1}{2}, M \to N', J', M')$ collisions in a magnetic field parallel to the initial wave vector, $\mathscr{H} \parallel {\bf k}$. The red solid lines show the $M'$-averaged cross sections for the $(N=0, J=\frac{1}{2} \to N', J')$ collisions.}
\label{fig:CaH_int_par}
\end{figure*}

\begin{figure*}[htbp]
\includegraphics[width=8cm]{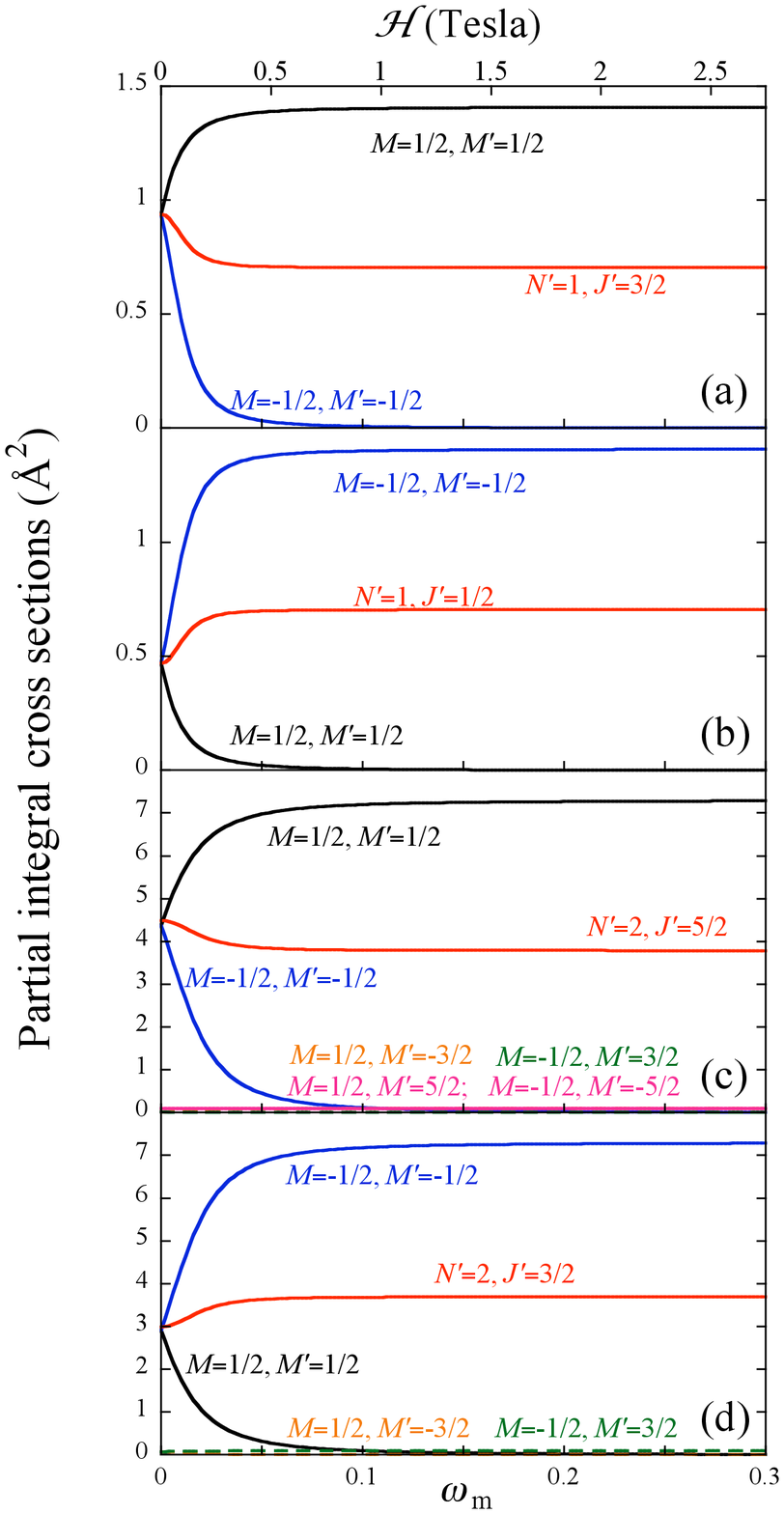}
\caption{Partial integral cross sections for the He -- CaH $(N=0, J=\frac{1}{2}, M \to N', J', M')$ collisions in a magnetic field perpendicular to the initial wave vector, $\mathscr{H} \perp {\bf k}$. The red solid lines show the $M'$-averaged cross sections for the $(N=0, J=\frac{1}{2} \to N', J')$ collisions.}
\label{fig:CaH_int_perp}
\end{figure*}

The integral cross-sections thus obtained for the magnetic field oriented parallel and perpendicular to the initial wave vector are presented in Figs.~\ref{fig:CaH_int_par} and~\ref{fig:CaH_int_perp}.  A prominent feature of the cross sections for the $N=0,J=\frac{1}{2} \to N', J'$ transitions is that, in the parallel field geometry, they increase for the $F_1$ final states and decrease for the $F_2$ states, while it is the other way around for the perpendicular geometry. 

In order to make sense of these trends in the field dependence of the $M$-averaged cross sections, let us take a closer look at the partial, $M$-resolved cross sections for the $N=0,J=\frac{1}{2},M \to N', J',M'$ channels and the two field geometries, also shown in Figs.~\ref{fig:CaH_int_par} and~\ref{fig:CaH_int_perp}. 
\\ \\
(i) Magnetic field \textit{parallel} to the initial wave vector, $\mathscr{H} \parallel {\bf k}$. In this case, the real Wigner matrices reduce to the Kronecker delta functions, $d_{-\rho, \Delta M}^{N'} (0) = \delta_{-\rho, \Delta M}$, and the scattering amplitude~(\ref{2SigScatAmpl}) becomes:
\begin{multline}
	\label{2SigScatAmplPar}	
	f_{0,\tfrac{1}{2},\pm\tfrac{1}{2} \to N',J',M'}^{\omega_\text{m}, \parallel} (\vartheta)= \frac{i k R_0}{2 \pi} \frac{\Xi_{N' 0} }{2N'+1} F_{N', -\Delta M} J_{\vert \Delta M \vert} (k R_0 \vartheta)  \\
	\times  \Biggl [ \pm a'(\omega_\text{m}) \sqrt{N' \mp M' +\tfrac{1}{2} } + b'(\omega_\text{m}) \sqrt{N' \pm M' +\tfrac{1}{2} }	\Biggr ]
\end{multline}
Eq.~(\ref{2SigScatAmplPar}) allows to readily interpret the dependences presented in Fig.~\ref{fig:CaH_int_par}. First, we see that the  $F_{N', -\Delta M}$ coefficients, defined by eq.~(\ref{Flamnu}), lead to a selection rule, namely that the cross sections vanish for $N' + \Delta M$ odd. Therefore, the partial cross sections for such combinations of $N'$ and $\Delta M$ do not contribute anything to the trends seen in Fig. \ref{fig:CaH_int_par}
 that we wish to explain. Equally absent are contributions from the transitions leading to the $F_1$ states with $M=\pm J$, since these states exhibit no alignment, see Fig.~\ref{fig:CaH_cos2}~(a),~(c). 
 
 As we can see from Fig.~\ref{fig:CaH_int_par}, the field-dependence of the cross section for the $N=0, J=1/2 \to N', J'$ transitions is a result of a competition among the partial $M$-resolved cross sections. Therefore we need  to account for the relative magnitudes of the non-vanishing $M$-dependent cross sections. Let us do it for the scattering channel $N=0, J=\frac{1}{2}, M \to N'=2, J'=\frac{5}{2}, M'$. Substituting the coefficients from Table~\ref{table:ab_coefs} into Eq.~(\ref{2SigScatAmplPar}), we see that the term in the square brackets vanishes for $M=-\frac{1}{2},M'=-\frac{1}{2}$ and for $M=-\frac{1}{2},M'=\frac{3}{2}$, but equals $\sqrt{5}$ for $M=\frac{1}{2},M'=\frac{1}{2}$ and $M=\frac{1}{2},M'=-\frac{3}{2}$. In addition, taking into account that $F_{2, 2} > F_{2, 0}$, we see that the $M$-averaged cross section must go up with increasing field strength.

 More generally, the dependence of the cross sections on the magnetic field is contained in the two hybridization coefficients $a'(\omega_\text{m})$ and $b'(\omega_\text{m})$. In the field-free case, $a'(\omega_\text{m}) = 0$ and $b'(\omega_\text{m}) = 1$ for collisions leading to $F_1$ states, whereas $a'(\omega_\text{m}) = 1$, $b'(\omega_\text{m}) = 0$ for collisions that produce $F_2$ states. In a magnetic field, the $a'(\omega_\text{m})$ and $b'(\omega_\text{m})$ coefficients assume values ranging between $-1$ and $+1$. As a consequence,  the $a'(\omega_\text{m})$ and $b'(\omega_\text{m})$ coefficients  have the same signs for the $F_1$  states and opposite signs for the $F_2$ states. Clearly, then, for an $F_1$ state, $|a'(\omega_\text{m})|$ increases with the field strength, while $|b'(\omega_\text{m})|$ decreases. Hence the factor in the square brackets of Eq.~(\ref{2SigScatAmplPar}) increases with $\omega_\text{m}$ for $M=\frac{1}{2}$ and decreases for $M=-\frac{1}{2}$, because of the opposite sign of the $a'(\omega_\text{m})$ coefficient.This is reversed for the final $F_2$ states, i.e., the cross sections increase for $M=-\frac{1}{2}$, and decrease for $M=\frac{1}{2}$. \\ \\

(ii) Magnetic field \textit{perpendicular} to the initial wave vector, $\mathscr{H} \perp {\bf k}$. In this case, Eq.~(\ref{2SigScatAmpl}) takes the form:

\begin{multline}
	\label{2SigScatAmplPerp}	
	f_{0,\tfrac{1}{2},\pm\tfrac{1}{2} \to N',J',M'}^{\omega_\text{m}, \perp} (\vartheta)= \frac{i k R_0}{2 \pi} \frac{\Xi_{N' 0} }{2N'+1} \left \{ \underset{\rho+N' \text{even}} {\sum_{\rho}} d_{-\rho, \Delta M}^{N'} (\tfrac{\pi}{2}) F_{N' \rho} J_{\vert \rho \vert} (k R_0 \vartheta)  \right \} \\
	\times  \Biggl [ \pm a'(\omega_\text{m}) \sqrt{N' \mp M' +\tfrac{1}{2} } + b'(\omega_\text{m}) \sqrt{N' \pm M' +\tfrac{1}{2} }	\Biggr ] 
\end{multline}
which is more involved than for the parallel case, although the field-dependent coefficients $a'(\omega_{\text{m}})$ and $b'(\omega_{\text{m}})$ remain outside of the summation.  The difference between the parallel and perpendicular cases is related to the values of the real Wigner $d$-matrices mixed by eq.~(\ref{2SigScatAmplPerp}). For instance, an inspection of the coefficients from Table~\ref{table:ab_coefs}, along with the $d_{-\rho, \Delta M}^{N'}$  matrices, reveals that the $M$-averaged cross sections for the $N=0, J=\frac{1}{2} \to N'=2, J'=\frac{5}{2}$ transition (i.e., transition to an $F_1$ final state) will decrease with $\omega_\text{m}$, in contrast to the parallel case.

%-----------------------------------------------------------------

\subsubsection{Frontal-versus-lateral steric asymmetry}
\begin{figure*}[htbp]
\centering\includegraphics[width=7.5cm]{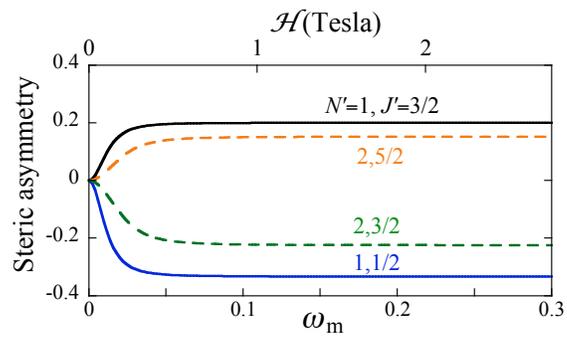}
\caption{Steric asymmetry, as defined by Eq.~(\ref{StericAsymmetry}), for the He -- CaH $(N=0, J=\frac{1}{2} \to N', J')$ collisions. Curves are labeled by $N', J'$.}
\label{fig:CaH_asym}
\end{figure*}

As in our previous work~\cite{LemFri1},~\cite{LemFri2}, we define a frontal-versus-lateral steric asymmetry by the expression
\begin{equation}
	\label{StericAsymmetry}
 	S_{\mathfrak{i} \to \mathfrak{f}} = \frac{\sigma_{\parallel}-\sigma_{\perp}}{\sigma_{\parallel}+\sigma_{\perp}}
\end{equation}
where the integral cross sections $\sigma_{\parallel, \perp}$ correspond, respectively, to $\mathscr{H} \parallel \mathbf{k}$ and $\mathscr{H} \perp \mathbf{k}$. The field dependence of the steric asymmetry for the He -- CaH collisions is presented in Fig.~\ref {fig:CaH_asym}. One can see that a particularly pronounced asymmetry obtains for the scattering into the $N'=1, J'=\frac{1}{2},\frac{3}{2}$ final states, while it is smaller for the $N=0 \to N'=2$ channels. Moreover, the steric asymmetry exhibits a sign alternation: it is positive for $F_1$ final states and negative for $F_2$ final states. This behavior is a reflection of the alternation in the trends of the integral cross sections for the $F_1$ and $F_2$ final states, discussed in the previous subsection, cf. the corresponding $M$-dependent integral cross sections, Figs.~\ref{fig:CaH_int_par} and~\ref{fig:CaH_int_perp}. 

We note that within the Fraunhofer model, elastic collisions do not exhibit any steric asymmetry. This follows from the isotropy of the elastic scattering amplitude, Eq. (\ref{AmplSphere}), which depends on the radius $R_0$ only: a sphere looks the same from any direction.

%--------------------------------------------------------------------------
%-------------------------- 3  S I G M A -------------------------------
%--------------------------------------------------------------------------

\section{Scattering of $^3\Sigma$ molecules by closed-shell atoms in a magnetic field}
\label{sec:3Sigma}

\subsection{A $^3\Sigma$ molecule in a magnetic field}
\label{sec:Zeem3Sigma}

The field-free Hamiltonian of a $^3 \Sigma$ electronic state consists of rotational, spin-rotation, and spin-spin terms
\begin{equation}
	\label{3SigFFhamil}
	H_0= B N^2 + \gamma \mathbf{N\cdot S} + \tfrac{2}{3} \lambda (3 S_z^2-S^2)
\end{equation}
where $\gamma$ and $\lambda$ are the spin-rotation and spin-spin constants, respectively. In the Hund's case~(b) basis, the field-free Hamiltonian~(\ref{3SigFFhamil}) consists of $3 \times 3$ matrices pertaining to different $J$ values (except for $J$=0). 
The matrix elements of Hamiltonian~(\ref{3SigFFhamil}) can be found, e.g., in ref.~\cite{amiot} (see also~\cite{BocaFri} and~\cite{Lefebvre-Brion}). The eigenenergies of $H_0$ are (in units of the rotational constant $B$):

\begin{equation}
	\label{3SigFFenergies}
	\begin{array}{l}
	E_1(J)/B = J(J+1) + 1 - \dfrac{3\gamma'}{2} - \dfrac{\lambda'}{3} - X \\[20pt]
	E_2(J)/B = J(J+1) - \gamma' + \dfrac{2 \lambda'}{3} \\[20pt]
	E_3(J)/B = J(J+1) + 1 - \dfrac{3\gamma'}{2} - \dfrac{\lambda'}{3} + X	
	\end{array}
\end{equation}
with
\begin{equation*}
         \begin{array}{l}
	X \equiv \left[ J(J+1)(\gamma' - 2)^2 + \left( \dfrac{\gamma' + 2 \lambda' - 2}{2} \right)^2  \right]^{1/2} \\[20pt]
	\gamma' \equiv \gamma/B\\[20pt]
 \lambda' \equiv \lambda/B
	\end{array}
\end{equation*}
The eigenenergies~(\ref{3SigFFenergies}) correspond to the three ways of combining rotational and electronic spin angular momenta $\mathbf{N}$ and  $\mathbf{S}$ for $S=1$ into a total angular momentum  $\mathbf{J}$; the total angular momentum quantum number takes values $J=N+1$, $J=N$, and $J=N-1$ for states which are conventionally designated as $F_1$, $F_2$, and $F_3$, respectively. For the case when $N=1$, $J=0$, the sign of the $X$ term should be reversed~\cite{Herzberg}. The parity of the states is  $(-1)^N$.

The interaction of a $^3 \Sigma$ molecule with a magnetic field $\mathscr{H}$ is given by:
\begin{equation}
	\label{MagnDipPotSig3}
	V_{\text{m}} = S_Z \omega_\text{m} B,
\end{equation}
where
\begin{equation}
	\label{omegapar3}
	\omega_\text{m} \equiv \frac{g_S \mu_B \mathscr{H}}{B} 
\end{equation}
is a dimensionless parameter characterizing the strength of the Zeeman interaction, cf. Eq. (\ref{omegapar2}).

We evaluated the Zeeman effect in Hund's case (b) basis 
\begin{equation}
	\label{EigenfuncFF}
	\vert N, J, M \rangle = c_{NJ}^1 \vert J, 1, M \rangle + c_{NJ}^0 \vert J, 0, M \rangle + c_{NJ}^{-1} \vert J, -1, M \rangle
\end{equation}
using the matrix elements of the $S_Z$ operator given in Appendix~\ref{app:sz}. 

The Zeeman eigenfunctions are hybrids of the Hund's case (b) basis functions~(\ref{EigenfuncFF}):
\begin{equation}
	\label{ZeemFunc3Sig}
	\left | \tilde{N},\tilde{J}, M; \omega_\text{m} \right >  = \sum_{NJ} a_{NJ}^{\tilde{N}\tilde{J}} (\omega_\text{m}) \left | N, J, M \right >
\end{equation}
and are labeled by $\tilde{N}$ and $\tilde{J}$, which are the angular momentum quantum numbers of the field-free state that adiabatically correlates with a given state in the field. Since $V_{\text{m}}$ couples Hund's case~(b) states that differ in $N$ by $0$ or $\pm2$, the parity remains definite in the presence of a magnetic field, and is given by $(-1)^{\tilde{N}}$. However, the Zeeman matrix for a $^3\Sigma$ molecule is no longer finite, unlike the $2\times 2$ Zeeman matrix for a $^2\Sigma$ state.

Using the Hund's case (b) rather than Hund's case~(a) basis set makes it possible to directly relate the field-free states and the Zeeman states, via the hybridization coefficients $a_{NJ}^{\tilde{N}\tilde{J}}$.

\begin{figure*}[htbp]
\centering
\includegraphics[width=7cm]{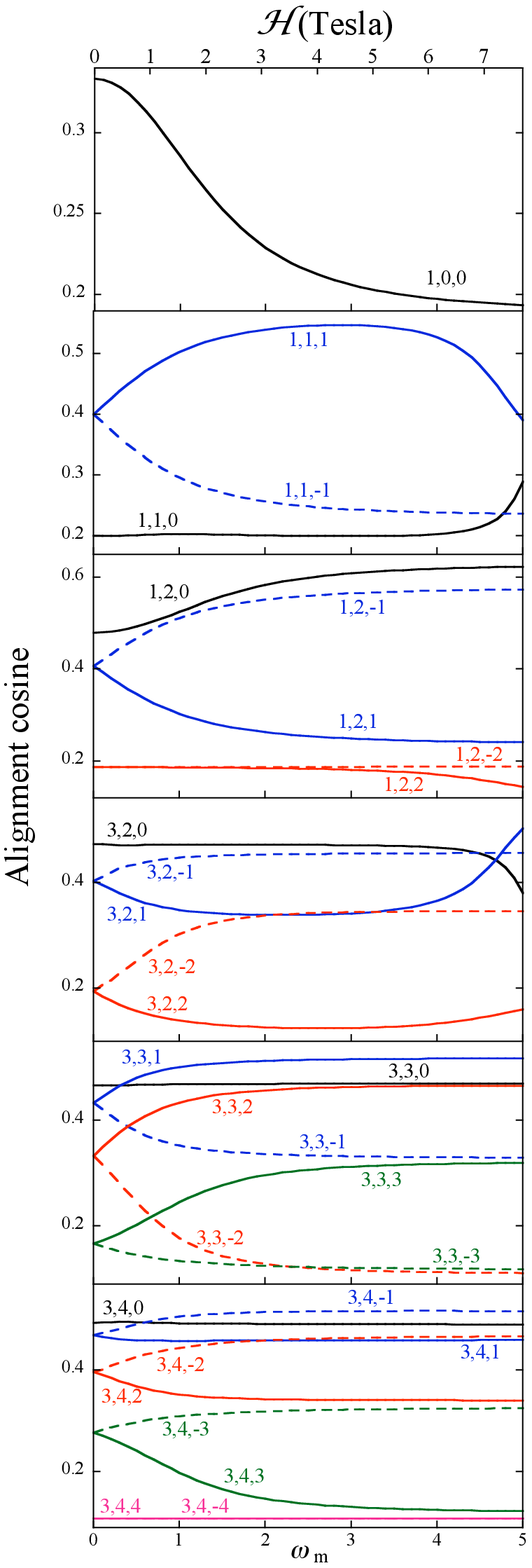}
\caption{Expectation values of the alignment cosine $\langle \cos^2\theta \rangle$ for the lowest Zeeman states of $^{16}$O$_2(X^3\Sigma^-)$ as a function of the magnetic field strength parameter $\omega_\text{m}$. States are labeled by $\tilde{N},\tilde{J},M$.}
\label{fig:O2_cos2}
\end{figure*}

The alignment cosine, $\langle \cos ^2 \theta \rangle$, of the Zeeman states can be evaluated from the matrix elements of Appendix~\ref{app:PhiZ2}. The dependence of $\langle \cos ^2 \theta \rangle$ on the magnetic field strength parameter $\omega_{\text{m}}$ is exemplified in Fig.~\ref{fig:O2_cos2} for $\tilde{N}=1,3$ Zeeman states of the $^{16}$O$_2$ molecule.

\subsection{The field-dependent scattering amplitude}
\label{sec:Fraun3Sigma}
We consider scattering from an initial $N, J$ state to a final $N', J'$ state. We transform the wavefunctions (\ref{ZeemFunc3Sig}) to the space-fixed frame by making use of Eq.~(\ref{FieldToSpaceDmatrix}) -- cf. Section \ref{sec:Fraun2Sigma}. As a result, the initial and final states become:

\begin{equation}
	\label{3SigInit}
	\vert \mathfrak{i}  \rangle \equiv \left | \tilde{N},\tilde{J}, M, \omega_\text{m} \right > = \sum_{NJ} \sqrt{\frac{2J+1}{4 \pi}} a_{NJ}^{\tilde{N}\tilde{J}} (\omega_\text{m}) \sum_{\Omega} c_{NJ}^{\Omega} \sum_{\xi} \mathscr{D}_{\xi M}^{J} (\varphi_{\varepsilon},\theta_{\varepsilon},0) \mathscr{D}_{\xi \Omega}^{J \ast} (\varphi, \theta,0)
\end{equation}

\begin{equation}
	\label{3SigFinal}
	\langle \mathfrak{f}  \vert \equiv \langle \tilde{N}',\tilde{J}', M', \omega_\text{m} \vert = \sum_{N'J'} \sqrt{\frac{2J'+1}{4 \pi}} b_{N'J'}^{\tilde{N}' \tilde{J}' } (\omega_\text{m}) \sum_{\Omega'} c_{N'J'}^{\Omega'} \sum_{\xi'} \mathscr{D}_{\xi' M'}^{J'} (\varphi_{\varepsilon},\theta_{\varepsilon},0) \mathscr{D}_{\xi' \Omega'}^{J' \ast} (\varphi, \theta,0)
\end{equation}
On substituting from Eqs.~(\ref{3SigInit}) and~(\ref{3SigFinal}) into Eq.~(\ref{InelAmplExpress}) and some angular momentum algebra, we obtain a general expression for the scattering amplitude:
\begin{multline}
	\label{3SigmaScatAmpl}
	f_{\mathfrak{i} \to \mathfrak{f}}^{\omega_\text{m}} (\vartheta)= \frac{i k R_0}{4 \pi} \underset{\kappa+\rho~\textrm{even}}{\underset{\kappa \neq 0 } {\sum_{\kappa \rho}}} \Xi_{\kappa 0} \mathscr{D}_{-\rho, \Delta M}^{\kappa \ast}  (\varphi_{\varepsilon},\theta_{\varepsilon},0)  F_{\kappa \rho} J_{\vert \rho \vert} (k R_0 \vartheta) \\
	\times  \underset{N'J'}{\sum_{NJ}} \sqrt{\frac{2J+1}{2J'+1}} a_{NJ}^{\tilde{N}\tilde{J}} (\omega_\text{m}) b_{N'J'}^{\tilde{N}' \tilde{J}' } (\omega_\text{m}) C(J \kappa J'; M \Delta M M') \sum_{\Omega} c_{NJ}^{\Omega} c_{N'J'}^{\Omega} C(J \kappa J'; \Omega 0 \Omega)
\end{multline}

%----------------------------------------------------------------------------------------------------------
%-------------------------   R E S U L T S     F O R    He - O2   --------------------------------
%----------------------------------------------------------------------------------------------------------

\subsection{Results for $\text{He--O}_2(X ^3\Sigma, N=0, J=1 \to N', J')$ scattering in a magnetic field}
\label{He-O2}
The $^{16}$O$_2(^3\Sigma^-)$ molecule has a rotational constant $B=1.4377$ cm$^{-1}$, a spin-rotation constant $\gamma=-0.0084$ cm$^{-1}$, and a spin-spin constant $\lambda=1.9848$ cm$^{-1}$~\cite{TomutaO2}. According to Ref.~\cite{PESHe-O2}, the ground state He -- O$_2$ potential energy surface has a global minimum of $-27.90$ cm$^{-1}$, which can be neglected at a collision energy $200$ cm$^{-1}$ (corresponding to a wave number $k=6.49$ \r{A}$^{-1}$). A small value of the Massey parameter, $\xi \approx 0.1$, ensures the validity of the sudden approximation. The ``hard shell" of the potential energy surface at this collision energy is shown in Fig.~\ref{fig:PEScut}, and  the Legendre moments $\Xi_{\kappa 0}$, obtained from a fit to the potential energy surface of Ref.~\cite{PESHe-O2}, are listed in Table~\ref{table:legendre_coefs}.  Since the He -- O$_2$ potential is of D$_{\text{2h}}$ symmetry, only even Legendre moments are nonzero.

Furthermore, since the nuclear spin of $^{16}$O is zero and the electronic ground state antisymmetric (a $^3\Sigma_g^-$ state), only rotational states with an odd rotational quantum number $N$ are allowed.  We will assume that the O$_2$ molecule is initially in its rotational ground state, $\vert N = 1, J = 0, M = 0 \rangle $.

Expression (\ref{3SigmaScatAmpl}) for the scattering amplitude further simplifies for particular geometries. In what follows, we will consider two such geometries.

(i) Magnetic field \textit{parallel} to the initial wave vector, $\mathscr{H} \parallel {\bf k}$, in which case the scattering amplitude becomes:
\begin{multline}
	\label{3SigmaScatAmplPar}
	f_{1, 0, 0 \to N', J', M'}^{\omega_\text{m},\parallel} (\vartheta)=   \frac{i k R_0}{4 \pi} J_{\vert  M' \vert} (k R_0 \vartheta) \underset{\kappa~\textrm{even}} {\sum_{\kappa \neq 0}} \Xi_{\kappa 0} F_{\kappa M'}  \\
	\times  \underset{N'J'}{\sum_{NJ}} \sqrt{\frac{2J+1}{2J'+1}} a_{NJ}^{1 0} (\omega_\text{m}) b_{N'J'}^{\tilde{N}' \tilde{J}' } (\omega_\text{m}) C(J \kappa J'; 0 M' M') \sum_{\Omega} c_{NJ}^{\Omega} c_{N'J'}^{\Omega} C(J \kappa J'; \Omega 0 \Omega)
\end{multline}

(ii) Magnetic field \textit{perpendicular} to the initial wave vector, $\mathscr{H} \perp {\bf k}$, in which case Eq.~(\ref{3SigmaScatAmpl}) simplifies to:

\begin{multline}
	\label{3SigmaScatAmplPerp}
	f_{1, 0, 0 \to N', J', M'}^{\omega_\text{m},\perp} (\vartheta)= \frac{i k R_0}{4 \pi} \underset{\kappa \neq 0} {\sum_{\kappa,\rho~\textrm{even}}}  \Xi_{\kappa 0}~d_{-\rho, M'}^{\kappa}  (\tfrac{\pi}{2})  F_{\kappa \rho} J_{\vert \rho \vert} (k R_0 \vartheta) \\
	\times  \underset{N'J'}{\sum_{NJ}} \sqrt{\frac{2J+1}{2J'+1}} a_{NJ}^{1 0} (\omega_\text{m}) b_{N'J'}^{\tilde{N}' \tilde{J}' } (\omega_\text{m}) C(J \kappa J'; 0 M' M') \sum_{\Omega} c_{NJ}^{\Omega} c_{N'J'}^{\Omega} C(J \kappa J'; \Omega 0 \Omega)
\end{multline}
Eqs. (\ref{3SigmaScatAmplPar}) and (\ref{3SigmaScatAmplPerp}) imply that for either geometry, only partial cross sections for the $N=1, J=0, M=0 \to N', J', M'$ collisions with $M'$ even can contribute to the scattering. This is particularly easy to see in the $\mathscr{H} \parallel \mathbf{k}$ case, where the $F_{\kappa M'}$ coefficients vanish for $M'$ odd as $F_{\kappa \rho}$ vanishes for odd $\kappa + \rho$. In the  $\mathscr{H} \perp \mathbf{k}$ case, a summation over $\rho$ arises. Since for $\kappa$ even and $M'$ odd the real Wigner matrices obey the relation $d_{-\rho, M'}^{\kappa}\left( \frac{\pi}{2} \right) = - d_{\rho, M'}^{\kappa}\left( \frac{\pi}{2} \right)$, the sum over $\rho$ vanishes and so do the partial cross sections for $M'$ odd.

%----------------------------------------------------------------------------------------------------------

\subsubsection{Differential cross sections}

\begin{figure*}[htbp]
\centering
\includegraphics[width=8cm]{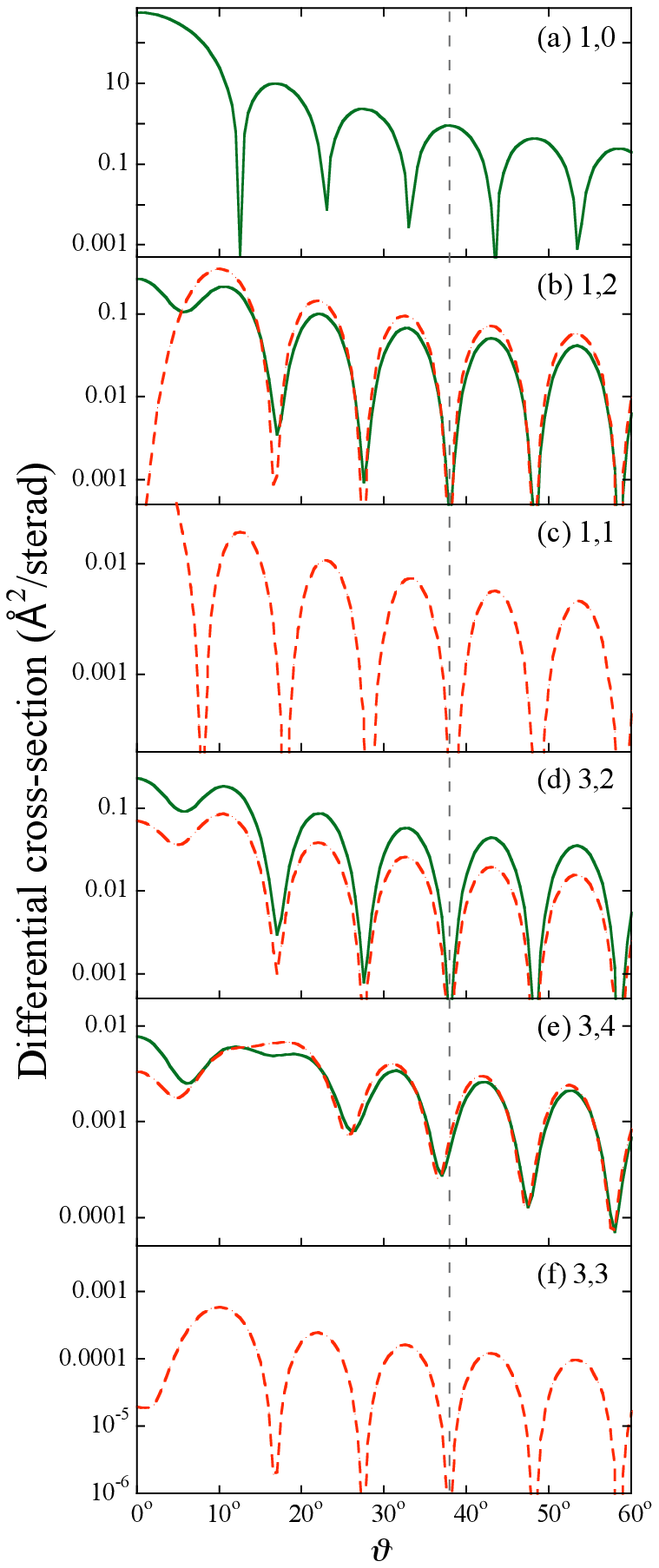}
\caption{Differential cross sections for the He -- O$_2$ $(N=1, J=0 \to N', J')$ collisions in a magnetic field $\omega_\text{m} = 5$ (red dashed line) parallel to the relative velocity vector, $\mathscr{H} \parallel {\bf k}$. The field-free cross sections are shown by the green solid line. The dashed vertical line serves to guide the eye in discerning the angular shifts of the partial cross sections. The field-free cross sections for the scattering to final states with $J'=N'$ vanish, see text.}
\label{fig:O2_diff_par}
\end{figure*}

\begin{figure*}[htbp]
\centering
\includegraphics[width=8cm]{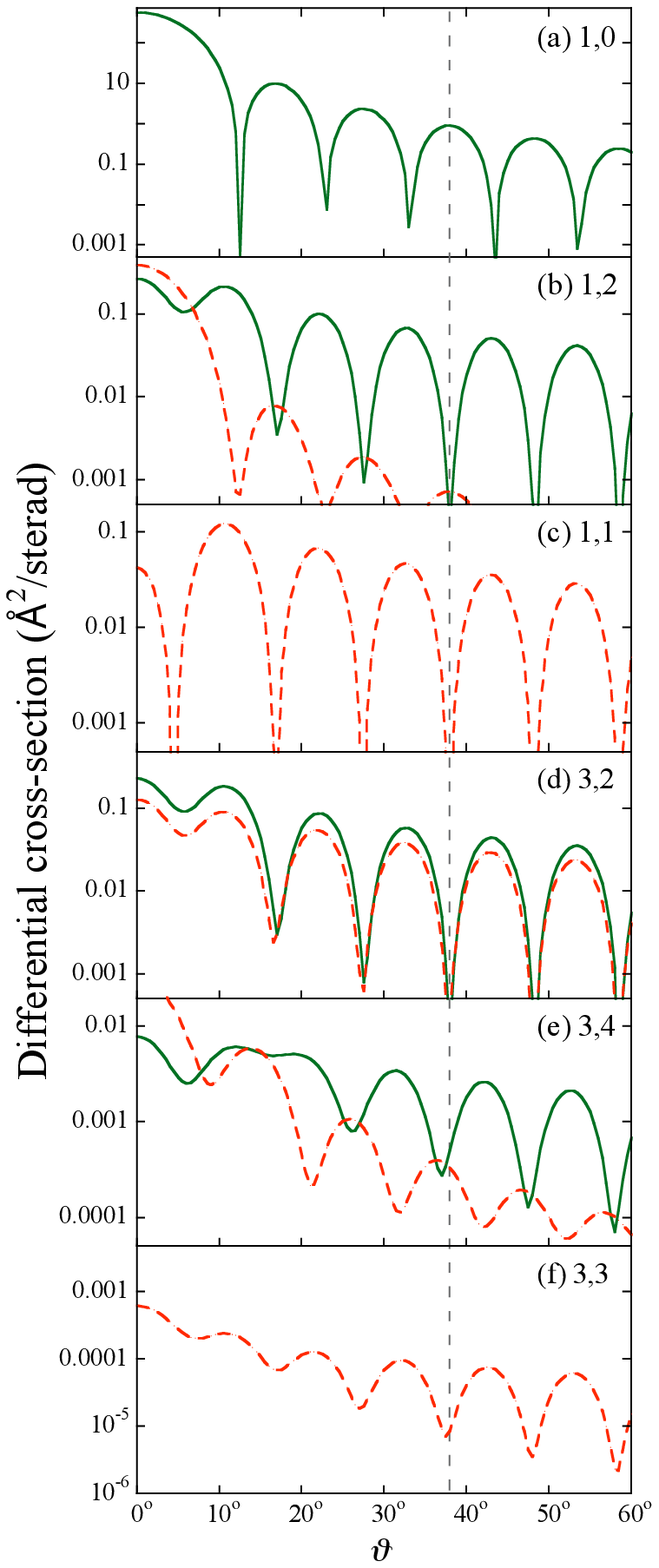}
\caption{Differential cross sections for the He -- O$_2$ $(N=1, J=0 \to N', J')$ collisions in a magnetic field $\omega_\text{m} = 5$ (red dashed line) perpendicular to the relative velocity vector, $\mathscr{H} \perp {\bf k}$. The field-free cross sections are shown by the green solid line. The dashed vertical line serves to guide the eye in discerning the angular shifts of the partial cross sections. The field-free cross sections for the scattering to final states with $J'=N'$ vanish, see text.}
\label{fig:O2_diff_perp}
\end{figure*}

The differential cross sections of the He -- O$_2$ $(N=1, J=0 \to N', J' )$ collisions, calculated from Eqs.~(\ref{DiffCrossFieldsJaver}) and~(\ref{DiffCrossFieldsJM}), are presented in Figs.~\ref{fig:O2_diff_par} and~\ref{fig:O2_diff_perp}. Also shown is the elastic cross section, obtained from the scattering amplitude~(\ref{AmplSphere}). The differential cross sections are shown for the field-free case, $\omega_\text{m}=0$, as well as for $\omega_\text{m}=5$, which for O$_2$ corresponds to a magnetic field  $\mathscr{H}$=7.7~T.

The angular dependence of the differential cross sections is determined by the Bessel functions appearing in the scattering amplitudes~(\ref{3SigmaScatAmplPar}) and~(\ref{3SigmaScatAmplPerp}). In the parallel case, the angular dependence is given expressly by $J_{\vert M' \vert} (k R_0 \vartheta)$, and is not affected by the magnetic field. Since only even-$\kappa$ terms contribute to the sum in the scattering amplitude and the coefficients $F_{\kappa \rho}$ vanish for $\kappa + \rho$ odd, the differential cross sections are given solely by even Bessel functions. This is the case for both parallel and perpendicular geometries. As the elastic scattering amplitude is given by an odd Bessel function, Eq. (\ref{AmplSphere}), the elastic and rotationally inelastic differential cross sections oscillate with an opposite phase.

 According to the general properties of Bessel functions~\cite{Watson}, at large angles the phase shift between different even Bessel functions disappears and their asymptotic form is given by Eq.~(\ref{AsymptBessel}). For the system under consideration, the phase shift between $J_0 (k R_0 \vartheta)$ and $J_2 (k R_0 \vartheta)$ functions becomes negligibly small at angles about 40$^\circ$, while the shift between $J_4 (k R_0 \vartheta)$ and either  $J_0 (k R_0 \vartheta)$ or $J_2 (k R_0 \vartheta)$ can only be neglected at angles about 120$^\circ$. Therefore, if the cross section is comprised only of $J_0$ and $J_2$ contributions, it will not be shifted with respect to the field-free case, while there may appear a field-iduced phase shift if the $J_4$ Bessel function also contributes. Indeed, in the parallel case, the $1, 0 \to 3, 4$ cross section, presented in Fig.~\ref{fig:O2_diff_par}~(e), exhibits a slight phase shift. In the perpendicular case, the Bessel functions $J_{\vert \rho \vert} (k R_0 \vartheta)$ for a range of $\rho$'s are mixed, see eq.~(\ref{3SigmaScatAmplPerp}), which results in a field-induced phase shift for both the $1, 0 \to 1, 2$ and $1, 0 \to 3, 4$ transitions. Figs.~\ref{fig:O2_diff_par} and~\ref{fig:O2_diff_perp} show the differential cross sections only up to 60$^\circ$, since this angular range dominates the integral cross sections. However, the phase shifts disappear only at larger angles, of about 120$^\circ$.
 
The most dramatic feature of the magnetic field dependence of the differential cross sections is the onset of inelastic scattering for channels that are closed in the absence of the field: these involve the  transitions $N=1, J=0 \to N', J' $ with $J'=N'$. That these channels are closed in the field-free case can be gleaned from the scattering amplitude~(\ref{3SigmaScatAmplPar}) for $\omega_\text{m}=0$, which reduces to
\begin{equation}
	\label{3SigmaScatAmplFF}
	f_{1, 0, 0 \to N', J', M'}^{FF} (\vartheta) = \frac{i k R_0}{4 \pi} J_{\vert  M' \vert} (k R_0 \vartheta)   \frac{ \Xi_{J' 0}}{\sqrt{2J'+1}} F_{J' M'} c_{N' J'}^{0}
\end{equation}
This field-free amplitude vanishes because the $c_{N' J'}^{0}$ coefficients are zero for $N'=J'$, as can be shown by the diagonalization of the field-free Hamiltonian~(\ref{3SigFFhamil}). As a result,  the field-free cross sections for the transitions to $1,1$ and $3,3$ states vanish. The hybridization by a magnetic field brings in coefficients $c_{N'J'}^{\pm 1}$ which are nonvanishing for $N'=J'$. The feature manifests itself in the integral cross sections as well.

%----------------------------------------------------------------------------------------------------------

\subsubsection{Integral cross sections}
\label{sec:O2intcross}

\begin{figure*}[htbp]
\centering
\includegraphics[width=8cm]{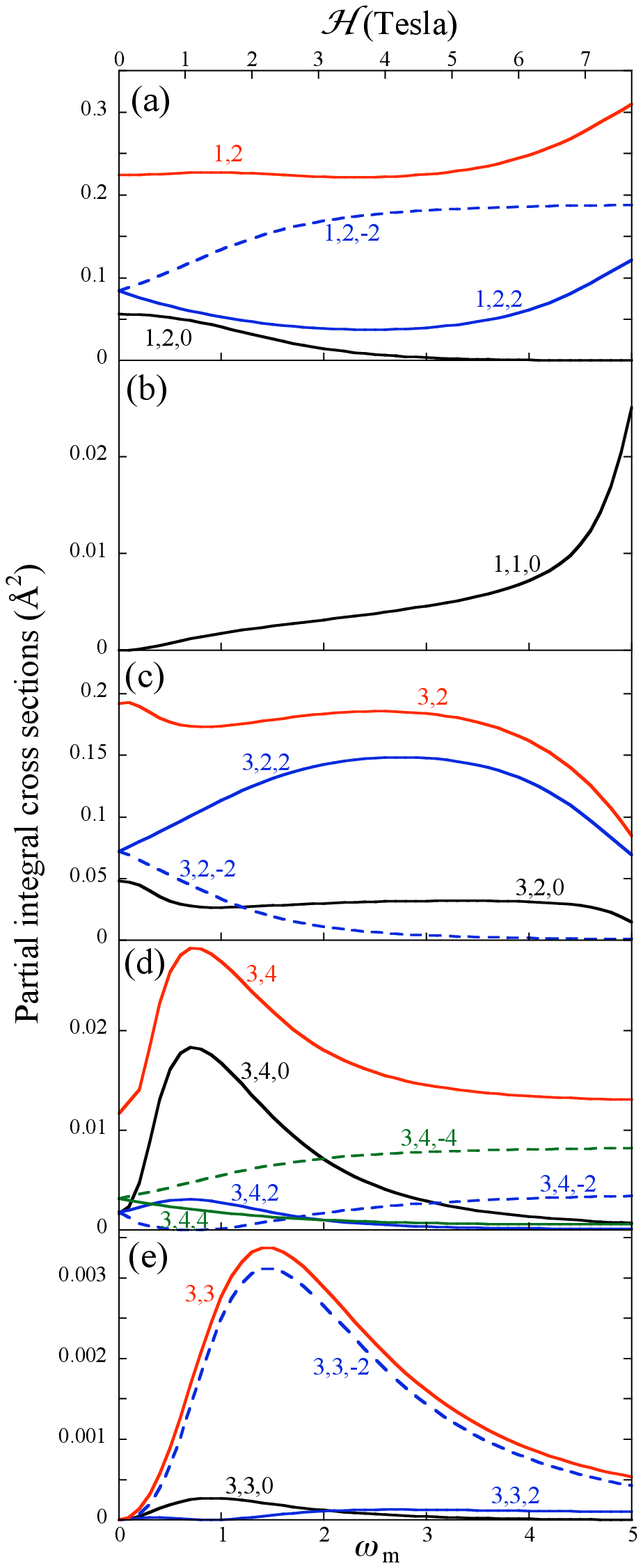}
\caption{Partial integral cross sections for the He -- O$_2$ $(N=1, J=0, M=0 \to N', J', M')$ collisions in a magnetic field parallel to the initial wave vector, $\mathscr{H} \parallel {\bf k}$. Curves are labeled by $N', J', M'$. The red solid lines show the $M'$-averaged cross sections. The partial cross sections, corresponding to negative $M'$, are shown by dashed lines.}
\label{fig:O2_int_par}
\end{figure*}

\begin{figure*}[htbp]
\centering
\includegraphics[width=8cm]{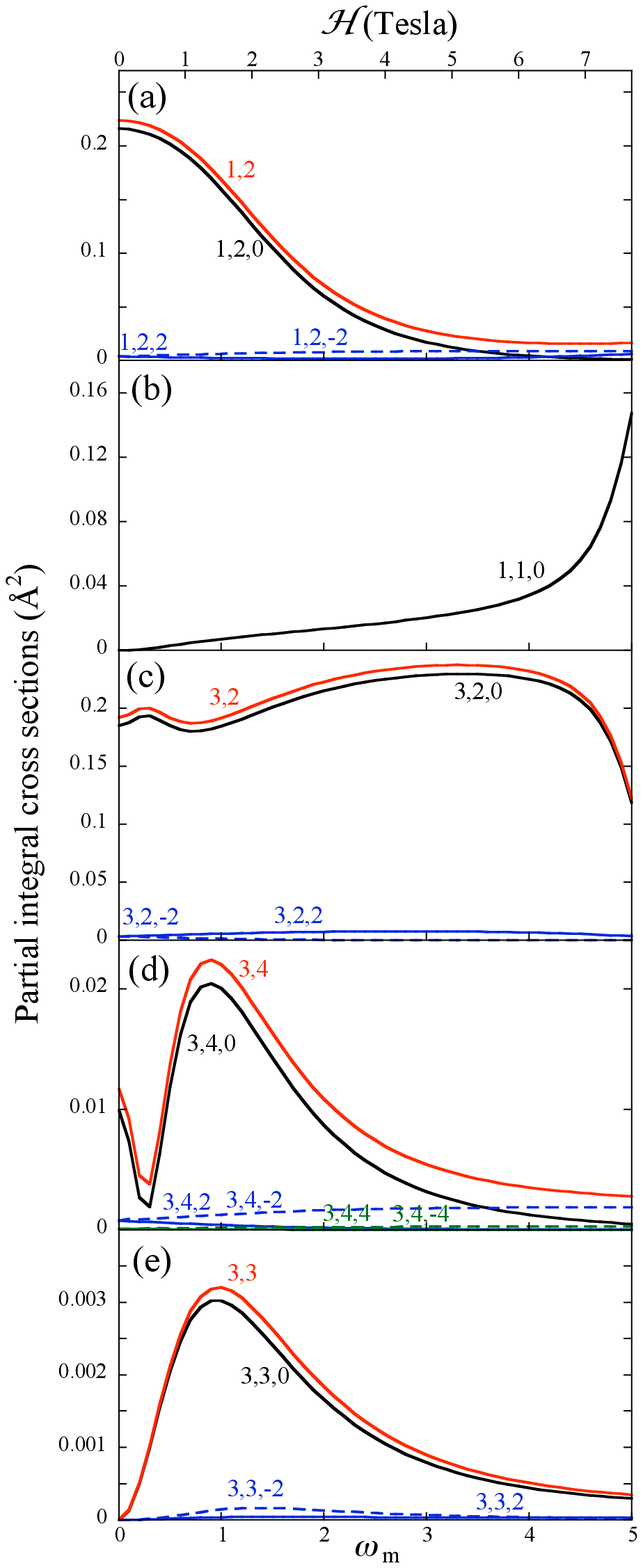}
\caption{Partial integral cross sections for the He -- O$_2$ $(N=1, J=0, M=0 \to N', J', M')$ collisions in a magnetic field perpendicular to the initial wave vector, $\mathscr{H} \perp {\bf k}$. Curves are labeled by $N', J', M'$. The red solid lines show the $M'$-averaged cross sections. The partial cross sections, corresponding to negative $M'$, are shown by dashed lines.}
\label{fig:O2_int_perp}
\end{figure*}

The integral cross sections for the He -- O$_2$ $(N=1, J=0, M=0 \to N', J', M')$ scattering are shown in Figs.~\ref{fig:O2_int_par} and~\ref{fig:O2_int_perp} for $\mathscr{H} \parallel \mathbf{k}$ and $\mathscr{H} \perp \mathbf{k}$, respectively. 

First we note that since the expansion of the He -- O$_2$ potential is dominated by $\Xi_{\kappa 0}$, see Table~\ref{table:legendre_coefs}, the sum in Eqs.~(\ref{3SigmaScatAmplPar}) and~(\ref{3SigmaScatAmplPerp}) over $\kappa$ can be approximated by the $\kappa=2$ term. In that case the Clebsch-Gordan coefficient $C(J,\kappa,J', 0 0 0)$ imposes a selection rule which limits the summation over $J$ and $J'$ to terms with $J'=J; J \pm 2$. In the field-free case, this selection rule is only satisfied for scattering from $J=0 \to J'=2$ and, therefore, only $N=1,J=0 \to N'=1, J'=2$ and $N=1,J=0 \to N'=3, J'=2$ cross sections can be expected to be sizable at zero field. Figs. \ref{fig:O2_int_par} and~\ref{fig:O2_int_perp} corroborate that this is indeed the case.

\begin{figure*}[htbp]
\centering\includegraphics[clip,width=7cm]{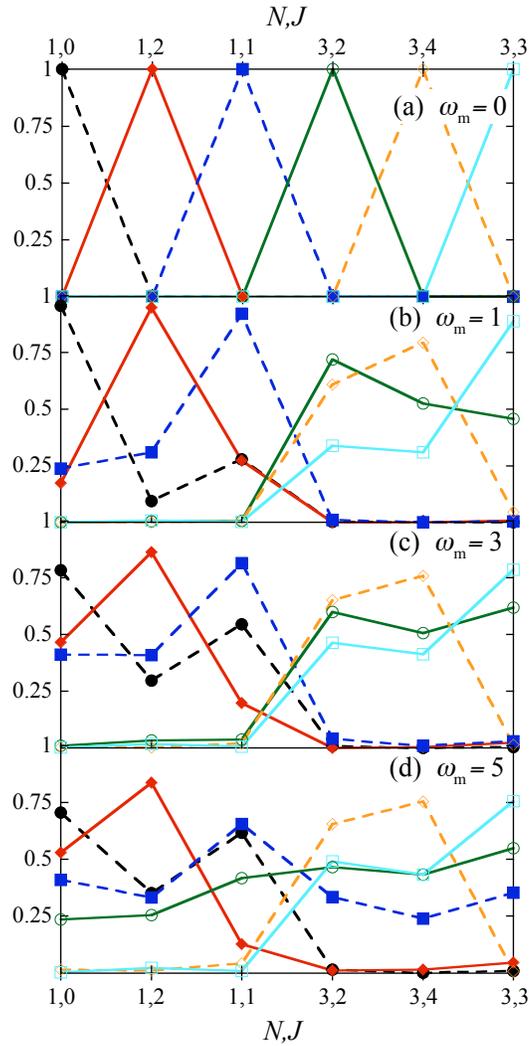}
\caption{Absolute values of the hybridization coefficients $a_{N J}^{1 0} (\omega_\text{m})$ (black dashed line, full circles) and $b_{N J}^{\tilde{N} \tilde{J}} (\omega_\text{m})$  for different values of the interaction parameter $\omega_\text{m}$, for the O$_2 (X^3 \Sigma^-)$ molecule. The following $b_{N J}^{\tilde{N} \tilde{J}} (\omega_\text{m})$ coefficients are presented:  $\tilde{N}=1, \tilde{J}=2$ (red solid line, full diamonds), $\tilde{N}=1, \tilde{J}=1$ (blue dashed line, full squares),$\tilde{N}=3, \tilde{J}=2$ (green solid line, empty circles), $\tilde{N}=3, \tilde{J}=4$ (orange dashed line, empty diamonds), $\tilde{N}=3, \tilde{J}=3$ (light blue solid line, empty squares). For all coefficients $M=0$. See text.}
\label{fig:O2_coefs_M0}
\end{figure*}

\begin{figure*}[htbp]
\centering\includegraphics[clip,width=7cm]{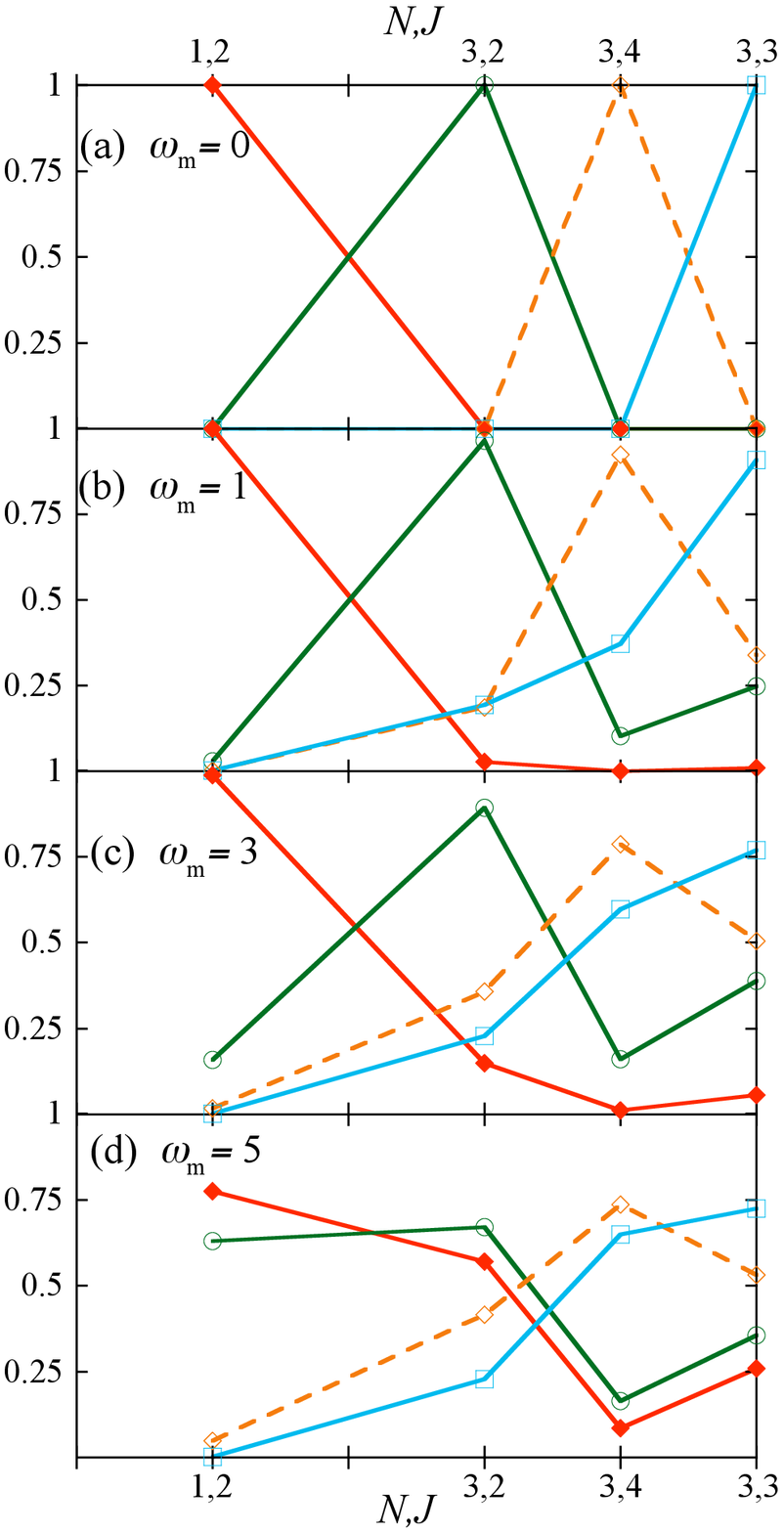}
\caption{Absolute values of the hybridization coefficients $b_{N J}^{\tilde{N} \tilde{J}} (\omega_\text{m})$  for different values of the interaction parameter $\omega_\text{m}$, for the O$_2 (X^3 \Sigma^-)$ molecule. The following $b_{N J}^{\tilde{N} \tilde{J}} (\omega_\text{m})$ coefficients are presented:  $\tilde{N}=1, \tilde{J}=2$ (red solid line, full diamonds), $\tilde{N}=3, \tilde{J}=2$ (green solid line, empty circles), $\tilde{N}=3, \tilde{J}=4$ (orange dashed line, empty diamonds), $\tilde{N}=3, \tilde{J}=3$ (light blue solid line, empty squares). For all coefficients $M=2$. See text.}
\label{fig:O2_coefs_M2}
\end{figure*}

\begin{figure*}[htbp]
\centering\includegraphics[clip,width=7cm]{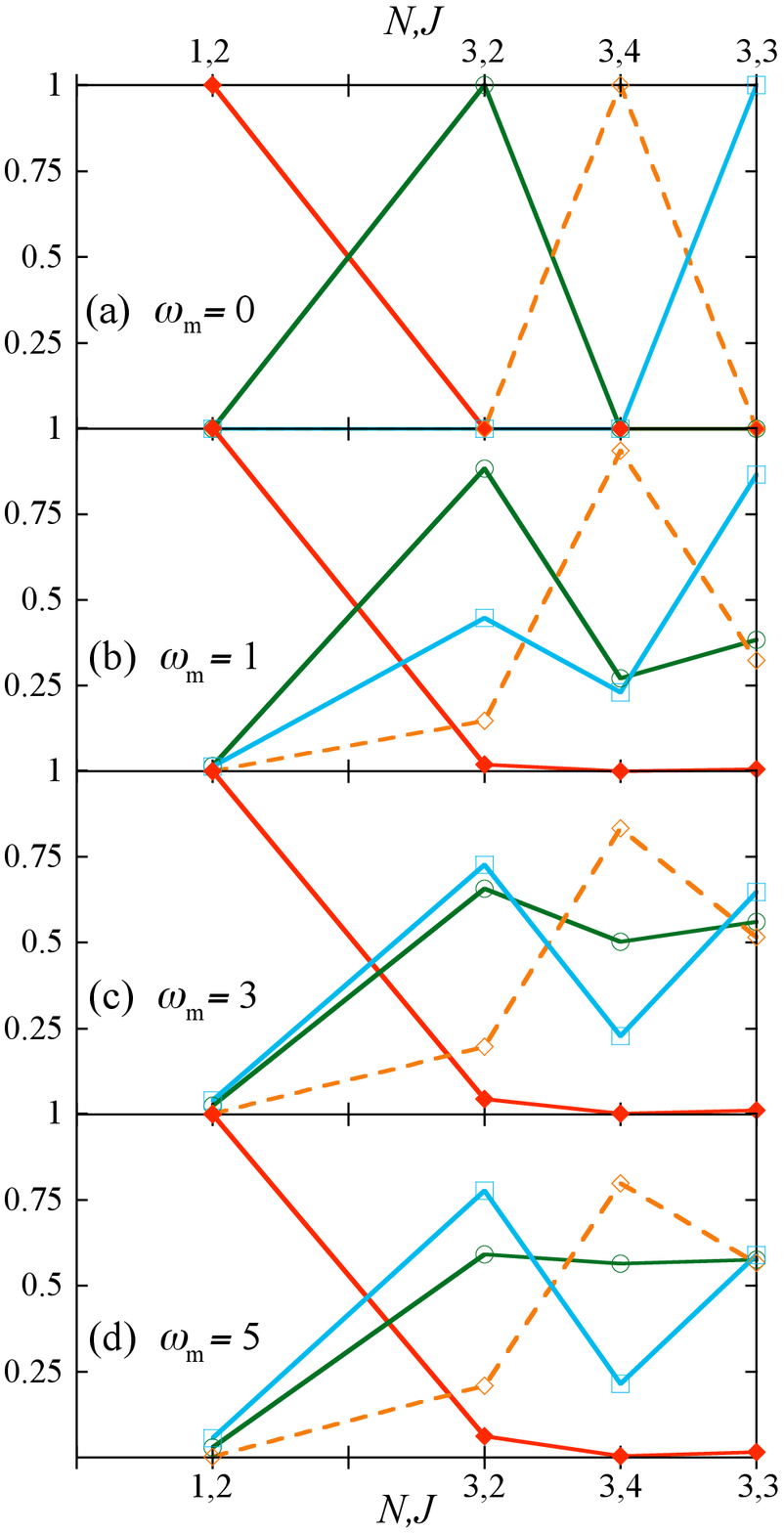}
\caption{Absolute values of the hybridization coefficients $b_{N J}^{\tilde{N} \tilde{J}} (\omega_\text{m})$  for different values of the interaction parameter $\omega_\text{m}$, for the O$_2 (X^3 \Sigma^-)$ molecule. The following $b_{N J}^{\tilde{N} \tilde{J}} (\omega_\text{m})$ coefficients are presented:  $\tilde{N}=1, \tilde{J}=2$ (red solid line, full diamonds), $\tilde{N}=3, \tilde{J}=2$ (green solid line, empty circles), $\tilde{N}=3, \tilde{J}=4$ (orange dashed line, empty diamonds), $\tilde{N}=3, \tilde{J}=3$ (light blue solid line, empty squares).  For all coefficients $M=-2$. See text.}
\label{fig:O2_coefs_Mmin2}
\end{figure*}

The field dependence of the $N=1, J=0 \to N', J'$ cross sections can be traced to the variation of the partial $N=1, J=0, M=0 \to N', J', M'$ contributions. Therefore, we will take a look at how the $M$-resolved  integral cross sections vary with the magnetic field. 
The field dependence of the partial $N=1, J=0, M=0 \to M', J', M'$ cross sections is contained in the hybridization coefficients $a_{NJ}^{1 0} (\omega_\text{m})$ and $b_{N'J'}^{\tilde{N}' \tilde{J}' } (\omega_\text{m})$, both for $\mathscr{H} \parallel \mathbf{k}$ and $\mathscr{H} \perp \mathbf{k}$.  The $a_{NJ}^{1 0}$ and $b_{N'J'}^{\tilde{N}' \tilde{J}' }$ coefficients for various values of $\tilde{N}', \tilde{J}'$ are presented in Figs.~\ref{fig:O2_coefs_M0},~\ref{fig:O2_coefs_M2}, and~\ref{fig:O2_coefs_Mmin2} for $M=0, 2$, and $-2$, respectively. At zero field, these coefficients are equal to unity for $N=\tilde{N}$, $J=\tilde{J}$ and are zero otherwise, see Figs.~\ref{fig:O2_coefs_M0}--\ref{fig:O2_coefs_Mmin2}~(a). Once the field is applied, the ``distributions" of the $a_{NJ}^{1 0} (\omega_\text{m})$ and $b_{N'J'}^{\tilde{N}' \tilde{J}' } (\omega_\text{m})$  coefficients undergo a broadening, as presented in Figs.~\ref{fig:O2_coefs_M0}--\ref{fig:O2_coefs_Mmin2}~(b)-(d). 

Such a broadening enhances the mixing of the $1,1,0$ state with the $1,2,0$ and $3,2,0$ states, leading to an increase of the $N=1,J=0 \to N'=1, J'=1$ cross section, see Figs.~\ref{fig:O2_int_par} and~\ref{fig:O2_int_perp}~(b). On the other hand, the broadening of the ``distribution" of the coefficients of the $1,2,0$ and $3,2,0$ final states reduces the overlap with the  initial state's coefficients and thus diminishes the $1,0,0 \to 1,2,0$ and $1,0,0 \to 3,2,0$ cross sections. 

The cross sections for the $3,3,0$ final state are distinctly non-monotonous, exhibiting maxima. These arise because at zero field the overlap of the hybridization coefficients for the $1,0,0$ and $3,3,0$ states is zero (if only $\kappa = 2$ is taken into account), but turning on the field  enhances the mixing of the $3,3,0$ and $3,2,0$ states, causing the corresponding cross section to increase. At higher $\omega_\text{m}$, the spread of coefficients becomes so large that the products $a_{NJ}^{1 0} b_{N'J'}^{3 3}$, corresponding to the selection rule $J'=J; J \pm 2$, become very small, cf.  Fig.~\ref{fig:O2_coefs_M0}~(d). As a result, the cross section for the $1,0,0 \to 3,3,0$ channel decreases again.

Figures~\ref{fig:O2_int_par} and~\ref{fig:O2_int_perp}~(a),(c)--(e) show that the cross sections corresponding to the final states with $M'=\pm 2$ exhibit maxima or minima, depending on the sign of $M'$. These are due to the changing overlap of the hybridization coefficients, just as for the  $3,3,0$ final state, see Figs.~\ref{fig:O2_coefs_M2} and~\ref{fig:O2_coefs_Mmin2}. 

Indeed, the overlaps of the coefficients corresponding to $M'=2$ and $M'=-2$ exhibit a different field dependence. For instance, the mixing of the $3,3,-2$ and $3,2,-2$ states, which increases with field strength, see Fig.~\ref{fig:O2_coefs_Mmin2}~(b), results in an increase of the $1,0,0 \to 1,2,-2$ cross section, presented in Fig.~\ref{fig:O2_int_par}~(a). For $M=2$, there is little mixing of the $3,4,2$ and $3,3,2$ states with the $3,2,2$ state, see Fig.~\ref{fig:O2_coefs_M2}~(b),~(c), which results in a decreasing $1,0,0 \to 1,2,2$ cross section. Once the coefficients' overlap increases at high fields, the corresponding cross section goes up again.

We note that both field-free and field-dressed transitions to the states with $M'=\pm 4$ are negligibly small, since their cross sections are dominated by other than the (dominant) $\Xi_{20}$ moment.  

The difference between the parallel and perpendicular geometries is due to the real $d$-matrices, appearing in Eq.~(\ref{3SigmaScatAmplPerp}). For instance, an inspection of equations~(\ref{3SigmaScatAmplPar}) and~(\ref{3SigmaScatAmplPerp}) reveals that the integral cross sections for the $1,0 \to 1,1$ transition will be larger for $\mathscr{H} \perp \mathbf{k}$ due to the coefficients $d^{\kappa}_{-\rho, 0}(\frac{\pi}{2})$. A similar argument holds in the case of scattering in electrostatic~\cite{LemFri1} and radiative~\cite{LemFri2} fields.

%----------------------------------------------------------------------------------------------------------

\subsubsection{Frontal-versus-lateral steric asymmetry}

\begin{figure*}[htbp]
\centering\includegraphics[width=7.5cm]{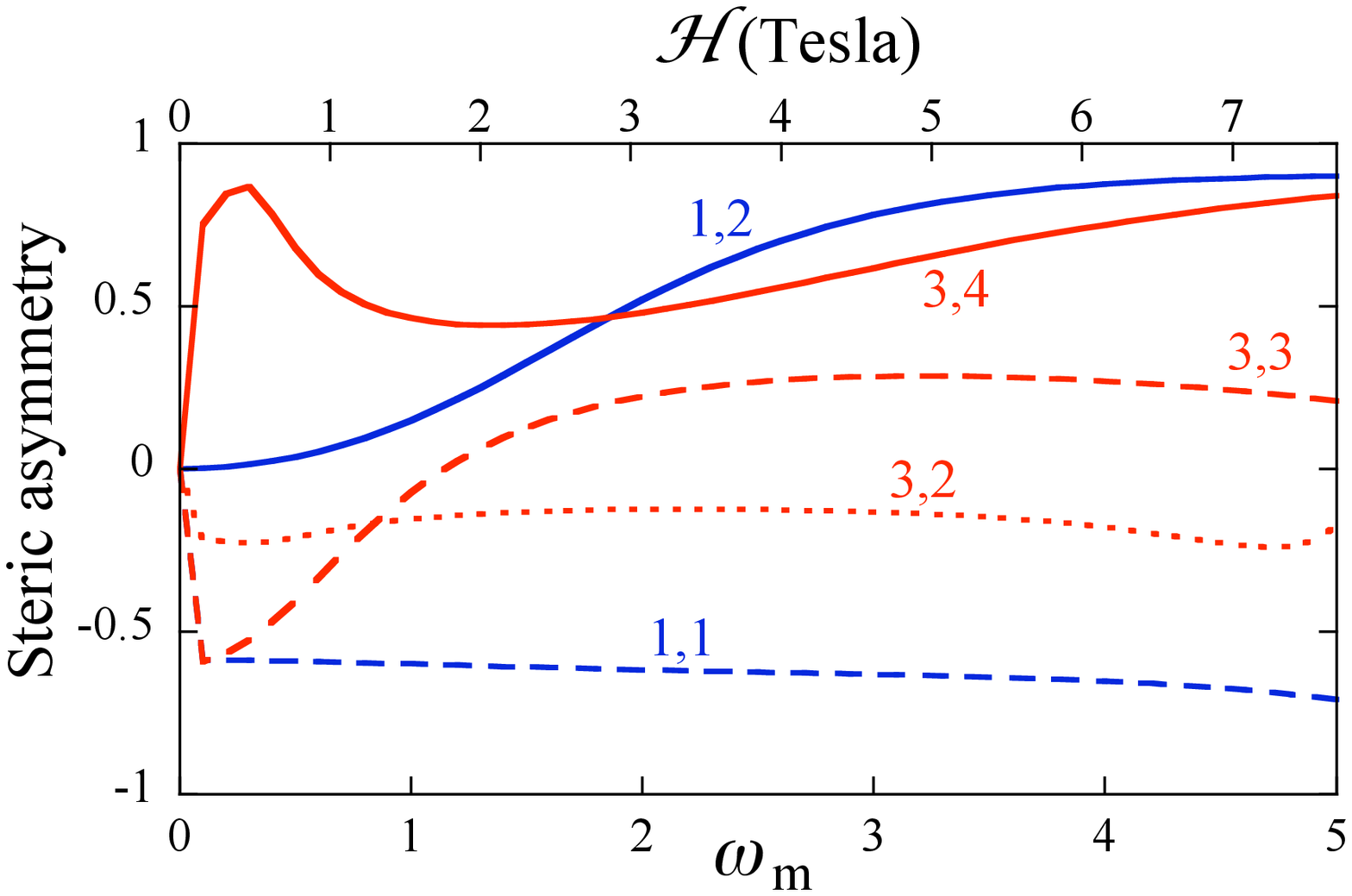}
\caption{Steric asymmetry, as defined by Eq.~(\ref{StericAsymmetry}), for the He -- O$_2$ $(N=1, J=0 \to N', J')$ collisions. Curves are labeled by $N', J'$.}
\label{fig:O2_asym}
\end{figure*}

Fig.~\ref{fig:O2_asym} shows the steric asymmetry for the He -- O$_2$ $(N=1, J=0 \to N', J')$ collisions as a function of the magnetic field. We see that the asymmetry is most pronounced for the $N=1, J=0 \to N=1, J=2$ and $N=1, J=0 \to N=3, J=4$ channels. This has its origin in the field dependence of the integral cross sections, see Figs.~\ref{fig:O2_int_par} and~\ref{fig:O2_int_perp}.

%----------------------------------------------------------------------------------------------------------
%-------------------------------  2 Pi     M O L E C U L E S   -------------------------------------
%----------------------------------------------------------------------------------------------------------

\section{Scattering of $^2\Pi$ molecules by closed-shell atoms in magnetic fields}
\label{sec:2Pi}

\subsection{The $^2\Pi$ molecule in magnetic field}
\label{sec:Zeem2Pi}

In this Section, we consider a Hund's case molecule, equivalent to a linear symmetric top. A good example of such a molecules is the $\text{OH}$ radical in its electronic ground state, $X ^2 \Pi_{\Omega}$, whose electronic spin and orbital angular momenta are strongly coupled to the molecular axis.
Each rotational state within the $^2 \Pi_{\Omega}$ ground state is equivalent to a symmetric-top state $\vert J, \Omega, M \rangle$ with projections $\Omega$ and $M$ of the total angular momentum $\mathbf{J}$ on the body- and space-fixed axes, respectively. Due to a coupling of the $\Pi$ state with a nearby $\Sigma$ state~\cite{BrownRot}, the levels with the same $\Omega$ are split into nearly-degenerate doublets whose members have opposite parities. The $\Omega$ doubling of the  $^2 \Pi_{\frac{3}{2}}$ state of OH increases as $J^3$, whereas that of the $^2 \Pi_{\frac{1}{2}}$ state increases linearly with $J$~\cite{Zare}. In our study, we used the values of the $\Omega$ doubling listed in Table~8.24 of ref.~\cite{BrownRot}. 

The definite-parity rotational states of a Hund's case (a) molecule can be written as
\begin{equation}
	\label{2PiSymmetrWF}
	\vert J, M, \Omega, \epsilon \rangle = \frac{1}{\sqrt{2}} \biggl [  \vert J, M,  \Omega  \rangle  + \epsilon \vert J, M,  - \Omega  \rangle \biggr ]
\end{equation}
where the symmetry index $\epsilon$ distinguishes between the members of a given $\Omega$ doublet.  Here and below we use the definition $\Omega \equiv |\Omega| $. The symmetry index takes the value of $+1$ or $-1$ for $e$ or $f$ levels, respectively. The parity of wave function (\ref{2PiSymmetrWF}) is equal to $\epsilon (-1)^{J-\frac{1}{2}}$~\cite{Brown75}. The rotational energy level structure of the OH radical in its $X ^2\Pi_{\Omega}$ state is reviewed in Sec.~2.1.4 of Ref.~\cite{Koos}.

%----------------------------------------------------------------------------------------------------------

When subject to a magnetic field, a Hund's case (a) molecule acquires a Zeeman potential  \begin{equation}
	\label{MagnDipPot2Pi}
	V_{\text{m}} = J_Z \omega_\text{m} B
\end{equation}
with $J_Z$ the $Z$ component of the total angular momentum (apart from nuclear spin), ${\mathbf J}$, and 
\begin{equation}
\label{omegapar2pi}
\omega_\text{m} \equiv (g_L\Lambda + g_S\Sigma) \mu_B \mathscr{H} / B
\end{equation}
Here $\Lambda$ and $\Sigma$ are projections of the orbital, $\mathbf{L}$, and spin, $\mathbf{S}$, electronic angular momenta on the molecular axis, $g_L=1$ and $g_S \simeq 2.0023$ are the electronic orbital and spin gyromagnetic ratios, $\mu_B$ is the Bohr magneton, $\mathscr{H}$ is the magnetic field strength, and $B$ is the rotational constant, cf. Eqs. (\ref{omegapar2}) and (\ref{omegapar3}). 
The matrix elements of Hamiltonian~(\ref{MagnDipPot2Pi}) in the definite-parity basis~(\ref{2PiSymmetrWF}) are
\begin{multline}
	\label{matrel2Pi}
	\langle J' M' \epsilon' \vert V_\text{m} \vert J M \epsilon \rangle = \omega_\text{m} B \left( \frac{1 + \epsilon \epsilon' (-1)^{J+J'+2\Omega}}{2} \right) (-1)^{J+J'+M-1/2} \\
	\times \sqrt{(2J+1) (2J'+1)}  \left( \begin{array}{ccc} j & 1 & j' \\  -\Omega & 0 & \Omega' \end{array} \right) \left( \begin{array}{ccc} j' & 1 & j \\  M & 0 & -M' \end{array} \right)
\end{multline}
where the last two factors are $3j$-symbols~\cite{Zare}, \cite{Varshalovich}. The matrix elements (\ref{matrel2Pi}) are a generalization of Eqs.~(\ref{SZoperator})--(\ref{PhiMinus}), and were presented, e.g., in Ref. \cite{Ticknor05}.
For an OH molecule in its ground $^2 \Pi_{3/2}$ state, the parity factor, $(1 + \epsilon \epsilon' (-1)^{J+J'+2\Omega})/2$, reduces to $\delta_{\epsilon \epsilon' }$, which means that the Zeeman interaction preserves parity. The Zeeman eigenstates are hybrids of the field-free states~(\ref{2PiSymmetrWF})
\begin{equation}
	\label{ZeemFunc2Pi}
	\left | \tilde{J}, M, \Omega, \epsilon; \omega_\text{m} \right >  = \sum_{J} a_{J M}^{\tilde{J}} (\omega_\text{m}) \vert J, M, \Omega, \epsilon \rangle
\end{equation}
where $\tilde{J}$ designates the angular momentum quantum number of the field-free state that adiabatically correlates with a given state in the field. The coefficients $a_{J M}^{\tilde{J}} (\omega_\text{m})$ can be obtained by the diagonalization of the Hamiltonian~(\ref{MagnDipPot2Pi}) in the basis~(\ref{2PiSymmetrWF}).

\begin{figure*}[htbp]
\centering
\includegraphics[width=7cm]{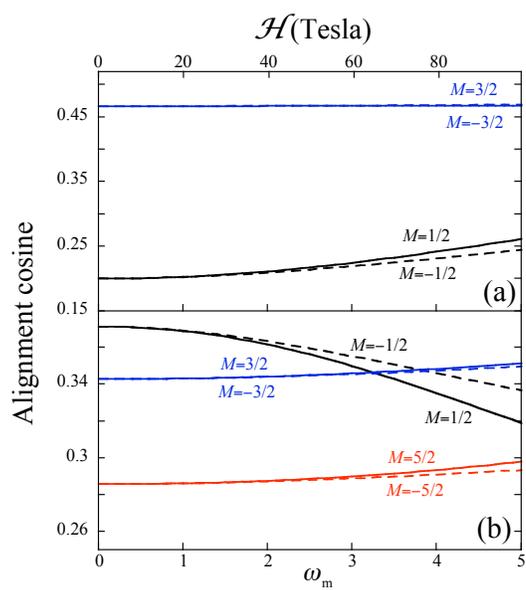}
\caption{Expectation values of the alignment cosine $\langle \cos^2\theta \rangle$ for the $3/2,f$ (a) and $5/2,f$ (b) states of the OH molecule, as a function of the magnetic field strength parameter $\omega_\text{m}$.}
\label{fig:OH_cos2}
\end{figure*}

The dependence of the alignment cosine, $\langle \cos ^2 \theta \rangle$, on the field strength parameter $\omega_{\text{m}}$ is shown in Fig.~\ref{fig:OH_cos2}  for the $3/2,f$ and $5/2,f$ states of the OH molecule. The matrix elements of the $\langle \cos ^2 \theta \rangle$ operator are listed in Appendix~\ref{app:PhiZ2}.

%----------------------------------------------------------------------------------------------------------

\subsection{The field-dependent scattering amplitude}
\label{sec:Fraun2Pi}

We consider scattering from the initial $J=3/2,e$ state to some $J', e/f$ state. As in the previous Sections, we use Eq.~(\ref{FieldToSpaceDmatrix}) to transform the wavefunctions~(\ref{2PiSymmetrWF}). Considering only the $\Omega$-conserving transitions ($\Omega' = \Omega$), the initial and final states are:
\begin{multline}
	\label{2PiInitial}
	\vert \mathfrak{i}  \rangle \equiv \left | \tilde{J}, M, \Omega, \epsilon, \omega_\text{m} \right > = \frac{1}{\sqrt{2}} \sum_{J} \sqrt{\frac{2J+1}{4 \pi}} a_{J M}^{\tilde{J}} (\omega_\text{m})  \sum_{\xi} \mathscr{D}_{\xi M}^{J} (\varphi_{\varepsilon},\theta_{\varepsilon},0) \left[  \mathscr{D}_{\xi \Omega}^{J \ast} (\varphi, \theta,0) + \epsilon  \mathscr{D}_{\xi -\Omega}^{J \ast} (\varphi, \theta,0) \right ]
\end{multline}
\begin{multline}
	\label{2PiFinal}
	\langle \mathfrak{f}  \vert \equiv \left < \tilde{J'}, M', \Omega, \epsilon', \omega_\text{m} \right | = \frac{1}{\sqrt{2}} \sum_{J'} \sqrt{\frac{2J'+1}{4 \pi}} b_{J' M'}^{\tilde{J'}} (\omega_\text{m})  \sum_{\xi'} \mathscr{D}_{\xi' M'}^{J' \ast} (\varphi_{\varepsilon},\theta_{\varepsilon},0) \left[  \mathscr{D}_{\xi' \Omega'}^{J'} (\varphi, \theta,0) + \epsilon'  \mathscr{D}_{\xi' -\Omega'}^{J'} (\varphi, \theta,0) \right ]
\end{multline}
By substituting Eqs.~(\ref{2PiInitial}) and~(\ref{2PiFinal}) into Eq.~(\ref{InelAmplExpress}), we obtain a closed expression for the scattering amplitude:
\begin{multline}
	\label{2PiScatAmpl}
	f_{\mathfrak{i} \to \mathfrak{f}}^{\omega_\text{m}} (\vartheta)= \frac{i k R_0}{4 \pi} \underset{\kappa+\rho~\textrm{even}}{\underset{\kappa \neq 0 } {\sum_{\kappa \rho}}} \Xi_{\kappa 0} \mathscr{D}_{-\rho, \Delta M}^{\kappa \ast}  (\varphi_{\varepsilon},\theta_{\varepsilon},0)  F_{\kappa \rho} J_{\vert \rho \vert} (k R_0 \vartheta) \\
	\times \sum_{J J'} \sqrt{\frac{2J+1}{2J'+1}} a_{J M}^{\tilde{J}} (\omega_\text{m}) b_{J' M'}^{\tilde{J'}} (\omega_\text{m}) C(J \kappa J'; M \Delta M M') C(J \kappa J'; \Omega 0 \Omega) \left [ 1 + \epsilon \epsilon' (-1)^{\kappa+\Delta J} \right]
\end{multline}
Eq.~(\ref{2PiScatAmpl}) simplifies for parallel or perpendicular orientations of the magnetic field with respect to the relative velocity vector. 

(i) For $\mathscr{H} \parallel \mathbf{k}$ we have

\begin{multline}
	\label{2PiScatAmplPar}
	f_{\mathfrak{i} \to \mathfrak{f}}^{\omega_\text{m},\parallel} (\vartheta)= \frac{i k R_0}{4 \pi} J_{\vert  \Delta M\vert} (k R_0 \vartheta)  \underset{\kappa+\Delta M~\textrm{even}} {\sum_{\kappa \neq 0 }} \Xi_{\kappa 0}F_{\kappa \Delta M} \\
	\times \sum_{J J'} \sqrt{\frac{2J+1}{2J'+1}} a_{J M}^{\tilde{J}} (\omega_\text{m}) b_{J' M'}^{\tilde{J'}} (\omega_\text{m}) C(J \kappa J'; M \Delta M M') C(J \kappa J'; \Omega 0 \Omega) \left [ 1 + \epsilon \epsilon' (-1)^{\kappa+\Delta J} \right]
\end{multline}

(ii) and for $\mathscr{H} \perp \mathbf{k}$ we obtain

\begin{multline}
	\label{2PiScatAmplPerp}
	f_{\mathfrak{i} \to \mathfrak{f}}^{\omega_\text{m}, \perp} (\vartheta)= \frac{i k R_0}{4 \pi} \underset{\kappa+\rho~\textrm{even}}{\underset{\kappa \neq 0 } {\sum_{\kappa \rho}}} \Xi_{\kappa 0} d_{-\rho, \Delta M}^{\kappa \ast}  (\tfrac{\pi}{2})  F_{\kappa \rho} J_{\vert \rho \vert} (k R_0 \vartheta) \\
	\times \sum_{J J'} \sqrt{\frac{2J+1}{2J'+1}} a_{J M}^{\tilde{J}} (\omega_\text{m}) b_{J' M'}^{\tilde{J'}} (\omega_\text{m}) C(J \kappa J'; M \Delta M M') C(J \kappa J'; \Omega 0 \Omega) \left [ 1 + \epsilon \epsilon' (-1)^{\kappa+\Delta J} \right]
\end{multline}

%----------------------------------------------------------------------------------------------------------
%-------------------------   R E S U L T S     F O R    He - OH   --------------------------------
%----------------------------------------------------------------------------------------------------------

\subsection{Results for $\text{He--OH}(X ^2\Pi_{\frac{3}{2}}, J=\frac{3}{2},f \to J',e/f)$ scattering in a magnetic field}
\label{He-OH}

According to Ref.~\cite{PESHe-OH}, the ground state He--OH potential energy surface has a global minimum of $-30.02$ cm$^{-1}$, which could be considered negligible with respect to a collision energy on the order of $100$ cm$^{-1}$, as for the He -- CaH and He -- O$_2$ systems treated above. However, the OH molecule has a large rotational constant, $B = 18.5348$~cm$^{-1}$, and so the Massey parameter~(\ref{MasseyParameter}) becomes significantly smaller than unity only at higher energies. Therefore, in order to ensure the validity of the sudden approximation, we had to work with a collision energy of 1000~cm$^{-1}$ ($k=13.86$~\r{A}$^{-1}$; Massey parameter $\xi \approx 0.5$). The corresponding equipotential line of the He -- OH ($^2 \Pi$) potential energy surface is shown in Fig.~\ref{fig:PEScut}, and the Legendre moments, $\Xi_{\kappa 0}$, obtained by fitting the surface, are listed in Table~\ref{table:legendre_coefs}.

Because of the negative spin-orbit constant, $A = - 139.051$~cm$^{-1}$~\cite{BrownRot}, the $\Omega$ doublet of the OH($X ^2\Pi_{\Omega})$ molecule is inverted, with the paramagnetic $^2 \Pi_{3/2}$ state as its ground state. Since $|A| \gg |B|$, we can see why the OH molecule can be well described by the Hund's case~(a) coupling scheme.

In what follows, we consider OH ($^2 \Pi$) radicals prepared in the $v=0, \Omega=\tfrac{3}{2}, J=\tfrac{3}{2},f$ state by hexapole state selection, like, e.g., in ref.~\cite{Beek00}. The molecules enter a magnetic field region where they collide with $^4$He atoms. The scattered molecules are state-sensitively detected in a field-free region.

%----------------------------------------------------------------------------------------------------------

\subsubsection{Differential cross sections}
\label{sec:OHDiffCrossSec}

\begin{figure*}[htbp]
\centering
\includegraphics[width=8cm]{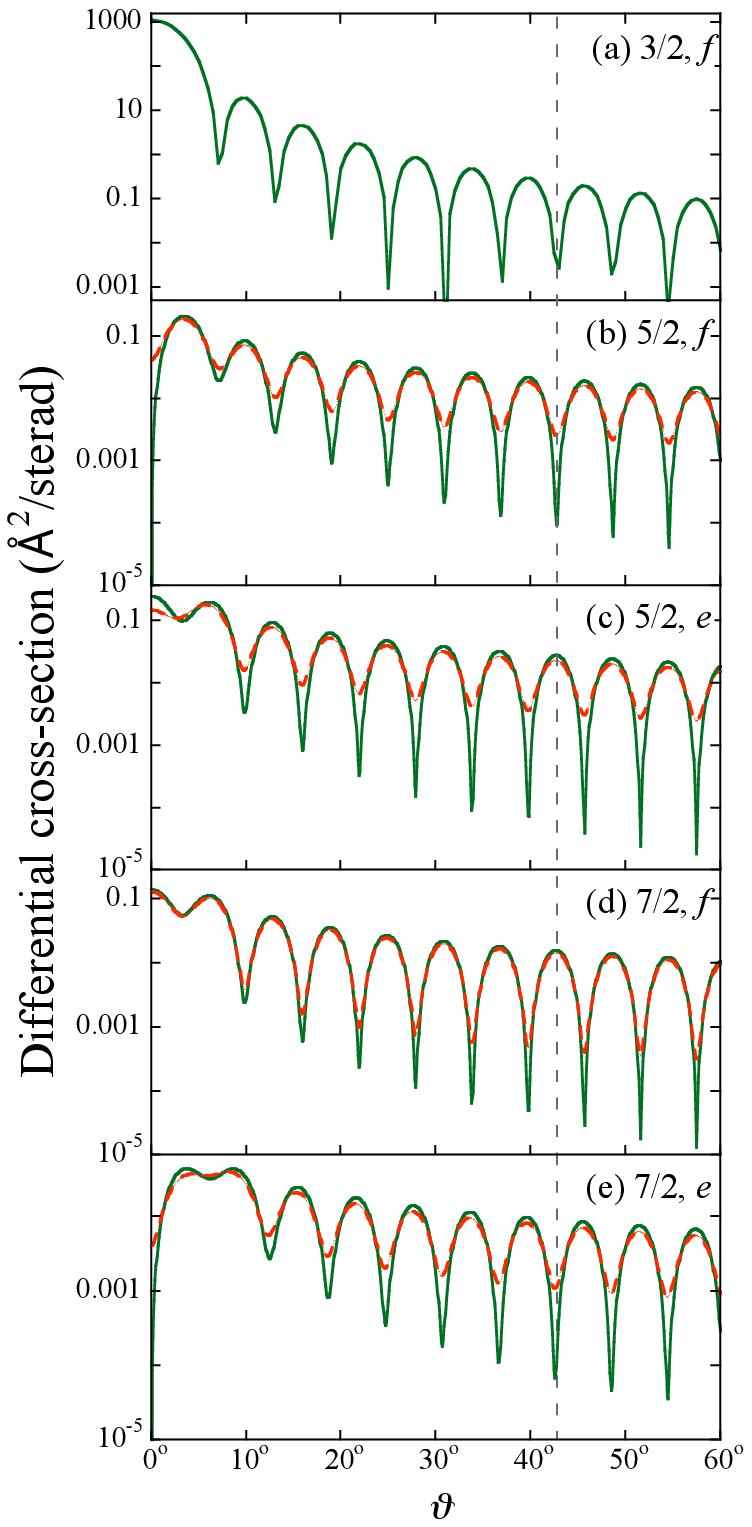}
\caption{Differential cross sections for the He -- OH $(J=3/2,f \to J',e/f)$ collisions in a magnetic field $\omega_\text{m} = 5$ (red dashed line) parallel to the relative velocity vector, $\mathscr{H} \parallel {\bf k}$. The field-free cross sections are shown by the green solid line. The dashed vertical line serves to guide the eye in discerning the angular shifts of the partial cross sections. See text.}
\label{fig:OH_diff_par}
\end{figure*}

\begin{figure*}[htbp]
\centering
\includegraphics[width=8cm]{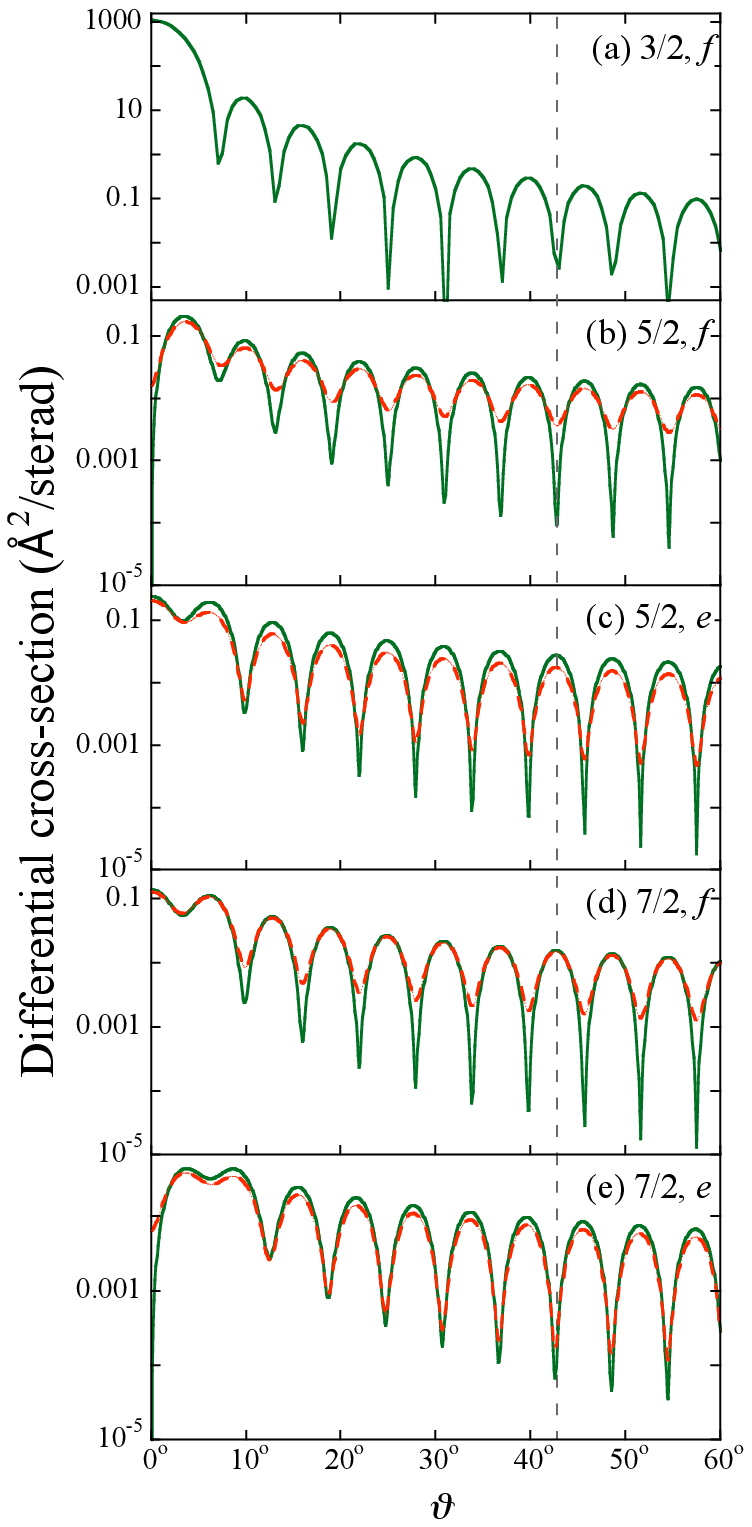}
\caption{Differential cross sections for the He -- OH $(J=3/2,f \to J',e/f)$ collisions in a magnetic field $\omega_\text{m} = 5$ (red dashed line) perpendicular to the relative velocity vector, $\mathscr{H} \perp {\bf k}$. The field-free cross sections are shown by the green solid line. The dashed vertical line serves to guide the eye in discerning the angular shifts of the partial cross sections. See text.}
\label{fig:OH_diff_perp}
\end{figure*}

We note that due to a large rotational constant, the Zeeman effect in the case of the OH radical is very weak, and so are the field-induced changes of the scattering. The differential cross sections for the He -- OH collisions, as obtained from Eqs.~(\ref{DiffCrossFieldsJaver}) and~(\ref{DiffCrossFieldsJM}), are shown in Figs.~\ref{fig:OH_diff_par} and~\ref{fig:OH_diff_perp}, together with the elastic scattering cross section obtained from Eq.~(\ref{AmplSphere}). The differential cross sections are presented for the field-free case, $\omega_\text{m}=0$, as well as for $\omega_\text{m}=5$, which for the OH radical corresponds to an extreme field strength of $\mathscr{H}$=99.2~T.  First, let us consider the field-free scattering amplitude
\begin{multline}
	\label{2PiScatAmplFF}	
	f_{\mathfrak{i} \to \mathfrak{f}}^{w=0} (\vartheta) =  
	\sqrt{\frac{2 J +1}{2J'+1}} J_{\vert \Delta M \vert} (k R_0 \vartheta) \\
	\times \underset{\kappa+\Delta M~\textrm{even}} {\sum_{\kappa \neq 0 }}  \Xi_{\kappa 0} F_{\kappa, \Delta M} C(J \kappa J'; M \Delta M M') C(J \kappa J'; \Omega 0 \Omega) \bigl[ 1 + \epsilon \epsilon' (-1)^{\kappa+\Delta J} \bigr ]
\end{multline}

We see that the angular dependence of the amplitude is given by the Bessel function $J_{\vert \Delta M \vert}$. The term in the square brackets and the  $F_{\kappa, \Delta M}$ coefficient provide a selection rule: $\Delta M + \Delta J$ must be even for parity conserving ($f \to f$) transitions, and odd for parity breaking ($f \to e$) transitions. The effect of this selection rule can be seen in Figs.~\ref{fig:OH_diff_par} and~\ref{fig:OH_diff_perp}. The elastic cross section, Figs.~\ref{fig:OH_diff_par}(a) and~\ref{fig:OH_diff_perp}(a), is proportional to an odd Bessel function, cf. Eq.~(\ref{AmplSphere}). Therefore, it is in phase with the $3/2,f \to 5/2,f$ and $3/2,f \to 7/2,e$ cross sections, but out of phase with $3/2,f \to 5/2,e$ and $3/2,f \to 7/2,f$ cross sections.

For a magnetic field parallel to the relative velocity, $\mathscr{H} \parallel \mathbf{k}$, the angular dependence is given explicitly by $J_{\vert \Delta M \vert} (k R_0 \vartheta)$, and is seen to be independent of the field, cf. Eq.~(\ref{2PiScatAmplPar}). Therefore, as Fig.~\ref{fig:OH_diff_par} shows, no field-induced phase shift of the differential cross sections takes place. For $\mathscr{H} \perp \mathbf{k}$, a mix of Bessel functions, $J_{\vert \rho \vert} (k R_0 \vartheta)$, contribute to the sum. However, since the Zeeman effect is so weak for the OH molecule, it is the $a_{J M}^{\tilde{J}} (\omega_\text{m})$, $b_{J' M'}^{\tilde{J'}} (\omega_\text{m})$ hybridization coefficients with $J = \tilde{J}$ which provide the main contribution to the sum, even at $\omega_\text{m}\approx 5$. As a result, no contributions from higher Bessel functions are drawn in, and so no significant field-induced phase shift is observed for the perpendicular case either.

%----------------------------------------------------------------------------------------------------------

\subsubsection{Integral cross sections}
\label{sec:OHIntCrossSec}
 
\begin{figure*}[htbp]
\centering
\includegraphics[width=8cm]{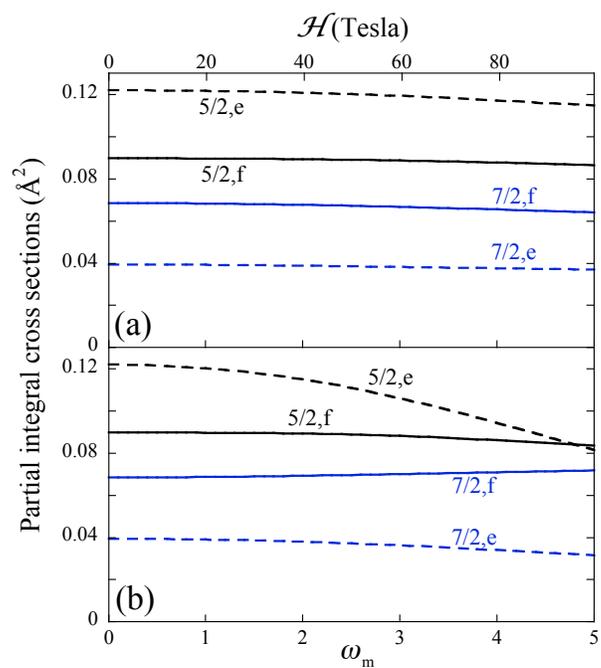}
\caption{Partial integral cross sections for the He -- OH $(J=3/2,f \to J',e/f)$ collisions in a magnetic field (a) parallel, $\mathscr{H} \parallel {\bf k}$, and (b) perpendicular, $\mathscr{H} \perp {\bf k}$, to the initial wave vector. Curves are labeled by $J',e/f$.}
\label{fig:OH_int}
\end{figure*}

\begin{figure*}[htbp]
\centering
\includegraphics[width=8cm]{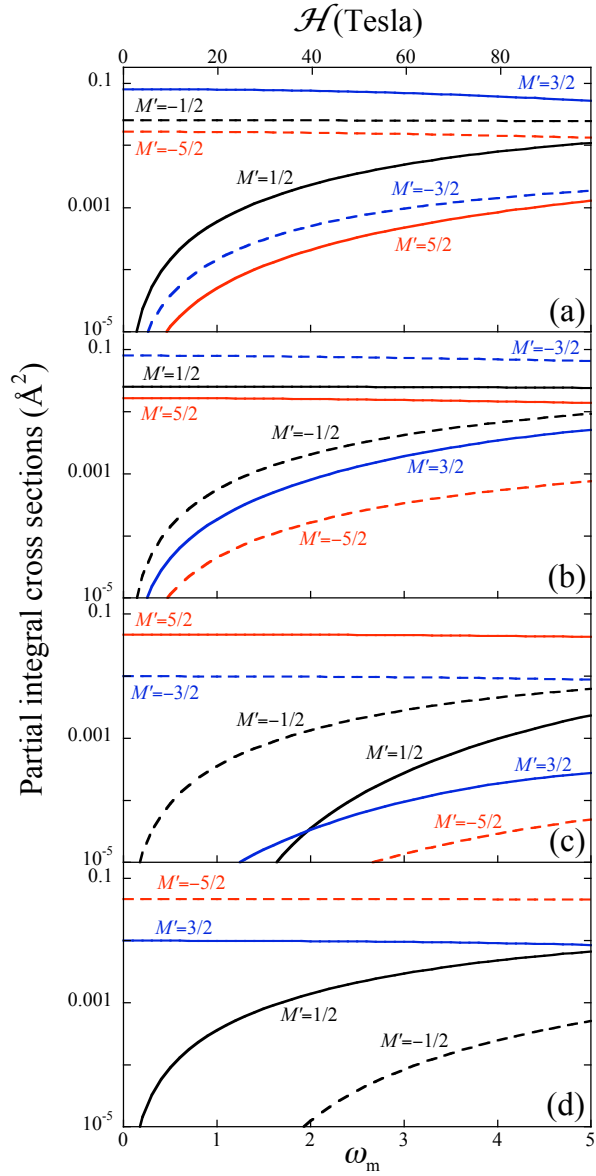}
\caption{Logarithm of the partial integral cross sections for the He -- OH $(J=3/2, f, M \to J'=5/2, f, M')$ collisions in a magnetic field parallel to the initial wave vector, $\mathscr{H} \parallel {\bf k}$. The panels correspond to different initial states: $M=1/2$~(a), $M=-1/2$~(b), $M=3/2$~(c), $M=-3/2$~(d).  All partial cross sections are non-vanishing, but the puny ones are not shown. See text.}
\label{fig:OH_int_m}
\end{figure*}

The integral cross sections for the He -- OH ($J=3/2,f \to J', e/f$) collisions are presented in Fig.~\ref{fig:OH_int} for $J'=5/2,7/2$ and the two orientations of the magnetic field with respect to the relative velocity. We see that no dramatic changes of the cross sections with the field strength take place. Again, the dependence of the $M$-averaged integral  cross sections (here $3/2,f \to J', e/f$) on the magnetic field can be traced to the field-dependences of the partial $M$- and $M'$-dependent contributions. Since for OH ($J=3/2,f $), the initial state comprises four $M$ values and the final state six or eight $M'$ values, a discussion of how  the $M$- and $M'$-averaged cross sections come about would be rather involved. Therefore, we resort to considering an example, namely how the $3/2,f \to 5/2,f$ cross section arises from the $3/2,f,M \to 5/2,f,M'$ components, shown in Fig.~\ref{fig:OH_int_m} for a magnetic field parallel to the relative velocity. From Table~\ref{table:legendre_coefs}, we see that the He -- OH potential is dominated by odd $\Xi_{\kappa 0}$ terms -- in contrast to the He -- CaH and He -- O$_2$ potentials. Therefore, odd $\Delta M$ values will yield the main contribution to the scattering amplitude~(\ref{2PiScatAmplPar}), because of the selection rule which dictates that $\kappa + \Delta M$ be even. This effect can be clearly discerned in Fig.~\ref{fig:OH_int_m}.
In the field-free case, even $\Delta M$ transitions have very small amplitudes, cf. Eq.~(\ref{2PiScatAmplFF}). When the field is on, the corresponding cross sections increase with $\omega_\text{m}$ due to the increasing overlap of the $a_{J M}^{\tilde{J}} (\omega_\text{m})$ and $b_{J' M'}^{\tilde{J'}} (\omega_\text{m})$ coefficients. This is a situation analogous to the one described in detail in Sec.~\ref{sec:O2intcross} for the O$_2$ -- He system (see also Figs.~\ref{fig:O2_coefs_M2},~\ref{fig:O2_coefs_Mmin2}).

%----------------------------------------------------------------------------------------------------------

\subsubsection{Frontal-versus-lateral steric asymmetry}
\label{sec:OHasymmetry}

\begin{figure*}[htbp]
\centering\includegraphics[width=7.5cm]{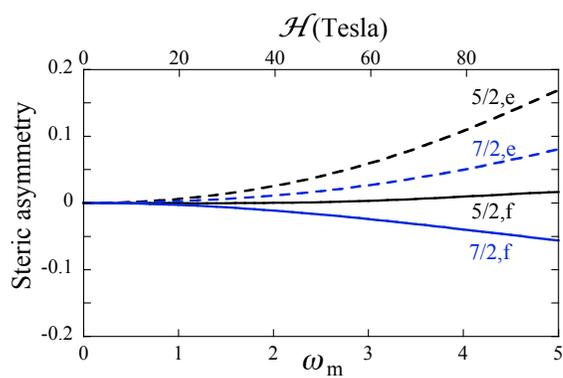}
\caption{Steric asymmetry, as defined by Eq.~(\ref{StericAsymmetry}), for the He -- OH $(J=3/2,f \to J',e/f)$ collisions. Curves are labeled by $J',e/f$.}
\label{fig:OH_asym}
\end{figure*}

The steric asymmetry for the He -- OH  ($J=3/2,f \to J', e/f$) collisions, calculated by means of Eq.~(\ref{StericAsymmetry}), is presented in Fig.~\ref{fig:OH_asym}. The most pronounced asymmetry is observed for the $5/2,e$ and $7/2,f$ channels, while the asymmetry for the $J=3/2,f \to 5/2, f$ and $J=3/2,f \to 7/2, e$ channels is  almost flat, especially at the feasible magnetic field strengths of up to 20 T. The difference between the scattering for parallel and perpendicular orientations of the magnetic field with respect to the initial velocity is contained in the $d_{-\rho, \Delta M}^{\kappa \ast}  (\tfrac{\pi}{2})$ matrices, appearing in Eq.~(\ref{2PiScatAmplPerp}), and the observed trends can be gleaned from Eqs.~(\ref{2PiScatAmplPar}) and~(\ref{2PiScatAmplPerp}).
%----------------------------------------------------------------------------------------------------------
%----------------------------------------------------------------------------------------------------------

\section{Conclusions}
\label{sec:conclusions}

We extended the Fraunhofer theory of matter wave scattering to tackle rotationally inelastic collisions of paramagnetic, open shell molecules with closed-shell atoms in magnetic fields. The description is inherently quantum and, therefore, capable of accounting for interference and other non-classical effects.  The effect of the magnetic field enters the model via the directional properties of the molecular states, which exhibit alignment of the molecular axis induced by the magnetic field. We applied the model to the  $\text{He--CaH}(X ^2\Sigma, J=1/2 \to J')$ and $\text{He--O}_2(X ^3\Sigma, N=0, J=1 \to N', J')$ scattering at a collision energy of 200 cm$^{-1}$, as well as to the $\text{He--OH}(X ^2\Pi, J=3/2,f \to J',e/f)$ collisions at an energy of 1000~cm$^{-1}$. 

In this Section, we mull over the results for the three collision systems and point out what they have in common and where they differ. 

The CaH molecule, studied in Sec.~\ref{sec:2Sigma}, has a non-magnetic $N=0,J=1$ ground state (taken as the initial state) and, therefore, all the field-induced changes of the He -- CaH scattering are due to the Zeeman effect of the final state. The magnetic eigenproperties of a $^2 \Sigma$ molecule can be obtained in closed form, by diagonalizing a $2 \times 2$ Hamiltonian matrix. When the magnetic field strength increases, the hybridization coefficients quickly approach an asymptotic value, see Table~\ref{table:ab_coefs}, as do the alignment cosine, Fig.~\ref{fig:CaH_cos2}, and the integral cross sections, Figs.~\ref{fig:CaH_int_par} and~\ref{fig:CaH_int_perp}. The changes of the cross sections between zero-field and the high-field limit are quite weak, which results in a small frontal-versus-lateral steric asymmetry of the He -- CaH scattering, Fig.~\ref{fig:CaH_asym}.

In  contrast to the $^2\Sigma$ case, the Hamiltonian matrices of the $^3 \Sigma$ and $^2 \Pi$ molecules in a magnetic field are in principle infinite. In practice, the Zeeman interaction couples a range of rotational states which is limited by the strength of the interaction to less than ten for $\omega_{\text m}\le 5$.

The He -- O$_2$ collision system exhibits a dramatic feature: in the absence of a magnetic field, the scattering vanishes for channels leading to the $F_2$ manifold, i.e., final states with $J'=N'$. However, in the presence of the field, such transitions become allowed, and, although weak, should be observable.  Also, some of the He -- O$_2$ integral cross sections are non-monotonous -- unlike the cross sections of the He -- CaH and He -- OH systems. They exhibit maxima or minima, which depend characteristically on the sign of $M'$, see Figs.~\ref{fig:O2_int_par} and~\ref{fig:O2_int_perp}. This contributes, for some channels, to the strong dependence of the He -- O$_2$ cross sections on the orientation of the magnetic field with respect to the relative velocity, as quantified by the frontal-versus-lateral steric asymmetry, Fig.~\ref{fig:O2_asym}.

The OH molecule has a large rotational constant and is, therefore, only weakly aligned by the magnetic field, see Fig.~\ref{fig:OH_cos2}. As a result, the field-induced changes of the scattering cross sections are tiny, Figs.~\ref{fig:OH_diff_par} -- \ref{fig:OH_int}, and so is the variation with field of the steric asymmetry, Fig.~\ref{fig:OH_asym}. Unlike the He -- CaH and He -- O$_2$ systems, the equipotential line on the He -- OH potential energy surface is dominated by odd Legendre moments, see Table~\ref{table:legendre_coefs}. This gives rise to scattering features which are qualitatively different from those of the other systems. For instance, as described in Sec.~\ref{sec:OHIntCrossSec}, it is the odd $\Delta M$ transitions that dominate the He -- OH $(J=3/2,f \to J', e/f)$ integral cross sections in a magnetic field parallel to the relative velocity vector. In the other two systems, it is the even $\Delta M$ transitions.

In all three systems studied, the field-induced changes of the differential cross sections -- such as angular shifts of their oscillations -- were puny. The only exception was found for the $N=1,J=0 \to N'=1,J'=2$ and $N=1,J=0 \to N'=3,J'=4$ transitions in the He -- O$_2$ system. These occur for scattering in a magnetic field perpendicular to the relative velocity vector, and are due to a field-induced mixing-in of higher Bessel-functions. 

The strength of the analytic model lies in its ability to separate dynamical and geometrical effects and to qualitatively explain the resulting scattering features. These include the angular oscillations in the state-to-state differential cross sections or the rotational-state dependent variation of the integral cross sections as functions of the magnetic field. We hope that the model will inspire new collisional experiments that make use either of crossed molecular beams or of a combination of a magnetic trap with a hot beam of atoms.

%========== A C K N O W L E D G E M E N T S ============================

\begin{acknowledgements}
Our special thanks are due to Gerard Meijer for discussions and support, and to Elena Dashevskaya and Evgueni Nikitin for their most helpful comments. We greatly enjoyed discussions with Bas van de Meerakker, Ludwig Scharfenberg,  Joop Gilijamse, and Steven Hoekstra. 
\end{acknowledgements}

%========== A P P E N D I C E S =======================================

\appendix

\section{Matrix elements of the $J_Z$ operator}
\label{app:sz}

In general, the Zeeman operator is proportional to the projection, $J_Z$, of the total electronic angular momentum, $\mathbf{J}$, on the space-fixed field axis, $Z$, see e.g. eq.~(\ref{MagnDipPot2Pi}). In this Appendix we present the matrix elements of the $J_Z$ operator, employed in this work. For $\Sigma$ electronic states, $J_Z$ reduces to $S_Z$. 

We transform the angular momentum projection operator from the body-fixed to the space-fixed coordinates using the direction cosines operator,~$\Phi$:
\begin{equation}
	\label{SZoperator}
	J_Z = \tfrac{1}{2} \left( \Phi_Z^{+} J^{-} + \Phi_Z^{-} J^{+} \right) + \Phi_Z^z J^z
\end{equation}
The matrix elements of the body-fixed spin operator  in the Hund's case (a) basis, $\vert J, \Omega, M \rangle$, are given by the standard relations~\cite{LLIII}:
\begin{align}
	\label{spmzoper}
	 \left <  J, \Omega, M \right |  J^{\pm} \left |  J, \Omega \mp 1, M \right > &=  \sqrt{(J \pm \Omega) (J \mp \Omega +1) } \\
	\left <  J, \Omega, M \right | J^z \left |  J, \Omega, M \right > &= \Omega
\end{align}

The matrix elements of the direction cosine operator can be obtained from Table 6 of~\cite{Hougen}. Some of them are also presented in Table 1.1 (p.19) of Ref.~\cite{Lefebvre-Brion}. We list here all non-vanishing matrix elements for $M'=M$:
\begin{equation}
	\label{Phiz}
	\langle J', \Omega, M \vert \Phi_{Z}^z \vert J, \Omega, M \rangle = \left \{ 
	\begin{array}{ll}
		\frac{\Omega M}{J (J+1)}  & \textrm{ } J' = J \\
		\\
		\frac{ \sqrt{(J+\Omega+1) (J-\Omega+1) (J+M+1) (J-M+1) } }{ (J+1) \sqrt{(2J+1) (2J+3)} }		 & \textrm{ } J' = J+1\\
		\\
		\frac{ \sqrt{(J+\Omega) (J-\Omega) (J+M) (J-M) } }{ J \sqrt{(2J+1) (2J-1)} } & \textrm{ } J' = J-1\\ 
	\end{array}    \right .
\end{equation}

\begin{equation}
	\label{PhiPlus}
	\langle J', \Omega-1, M \vert \Phi_{Z}^{+} \vert J, \Omega, M \rangle = \left \{ 
	\begin{array}{ll}
		\frac{M \sqrt{(J+\Omega) (J - \Omega+1) }}{ J(J+1)} & \textrm{ } J' = J \\
		\\
		\frac{\sqrt{ (J-\Omega+1) (J-\Omega+2) (J+M+1) (J-M+1) }}{ (J+1) \sqrt{(2J+1) (2J+3) } }  & \textrm{ } J' = J+1  \\
		\\
		- \frac{\sqrt{ (J+\Omega) (J+\Omega-1) (J+M) (J-M) }}{ J \sqrt{(2J+1) (2J-1) } } & \textrm{ } J' = J-1 \\
		\end{array}    \right .
\end{equation}

\begin{equation}
	\label{PhiMinus}
	\langle J', \Omega+1, M \vert \Phi_{Z}^{-} \vert J, \Omega, M \rangle = \left \{ 
	\begin{array}{ll}
		\frac{M \sqrt{(J-\Omega) (J + \Omega+1) }}{ J(J+1)} & \textrm{ } J' = J \\
		\\
		- \frac{\sqrt{ (J+\Omega+1) (J+\Omega+2) (J+M+1) (J-M+1) }}{ (J+1) \sqrt{(2J+1) (2J+3) } }  & \textrm{ } J' = J+1  \\
		\\
		\frac{\sqrt{ (J-\Omega) (J-\Omega-1) (J+M) (J-M) }}{ J \sqrt{(2J+1) (2J-1) } } & \textrm{ } J' = J-1 \\
		\end{array}    \right .
\end{equation}

\section{Matrix elements of the $(\Phi_Z^{z})^2 $ operator}
\label{app:PhiZ2}

The matrix elements of the alignment cosine can be reduced to the matrix elements of the $ \Phi_{Z}^z$ operator by means of the full basis set:
\begin{equation}
	\label{PhiZ2}
         \langle \cos^2 \theta \rangle = \langle J', \Omega, M \vert (\Phi_{Z}^{z})^2 \vert J, \Omega, M \rangle = \sum_{J''} \langle J', \Omega, M \vert \Phi_{Z}^{z} \vert J'', \Omega, M \rangle \langle J'', \Omega, M \vert \Phi_{Z}^{z} \vert J, \Omega, M \rangle
\end{equation}
By taking into account that the matrix elements~(\ref{Phiz}) are nonzero only for $\Delta J = 0, \pm 1$, we obtain the matrix elements of the direction cosine operator:
\begin{multline}
	\label{Phiz2JJ}
	\langle J, \Omega, M \vert (\Phi_{Z}^{z})^2 \vert J, \Omega, M \rangle =\left | \langle J-1, \Omega, M \vert \Phi_{Z}^z \vert J, \Omega, M \rangle \right |^2 +\left | \langle J, \Omega, M \vert \Phi_{Z}^z \vert J, \Omega, M \rangle \right |^2 \\
	+ \left | \langle J+1, \Omega, M \vert \Phi_{Z}^z \vert J, \Omega, M \rangle \right |^2  
\end{multline}
\begin{multline}
	\label{Phiz2JJm1}
	\langle J-1, \Omega, M \vert (\Phi_{Z}^{z})^2 \vert J, \Omega, M \rangle =
	\langle J-1, \Omega, M \vert \Phi_{Z}^z \vert J, \Omega, M \rangle \biggl \{  \langle J, \Omega, M \vert \Phi_{Z}^z \vert J, \Omega, M \rangle 
	+ \langle J-1, \Omega, M \vert \Phi_{Z}^z \vert J-1, \Omega, M \rangle \biggr \}
\end{multline}
\begin{multline}
	\label{Phiz2JJp1}
	\langle J+1, \Omega, M \vert (\Phi_{Z}^{z})^2 \vert J, \Omega, M \rangle =
	\langle J+1, \Omega, M \vert \Phi_{Z}^z \vert J, \Omega, M \rangle \biggl \{ \langle J, \Omega, M \vert \Phi_{Z}^z \vert J, \Omega, M \rangle 
	+ \langle J+1, \Omega, M \vert \Phi_{Z}^z \vert J+1, \Omega, M \rangle \biggr \}
\end{multline}
\begin{equation}
	\label{Phiz2JJm2}
	\langle J-2, \Omega, M \vert (\Phi_{Z}^{z})^2 \vert J, \Omega, M \rangle =
	\langle J-2, \Omega, M \vert \Phi_{Z}^z \vert J-1, \Omega, M \rangle \langle J-1, \Omega, M \vert \Phi_{Z}^z \vert J, \Omega, M \rangle
\end{equation}

\begin{equation}
	\label{Phiz2JJp2}
	\langle J+2, \Omega, M \vert (\Phi_{Z}^{z})^2 \vert J, \Omega, M \rangle =
	\langle J+2, \Omega, M \vert \Phi_{Z}^z \vert J+1, \Omega, M \rangle \langle J+1, \Omega, M \vert \Phi_{Z}^z \vert J, \Omega, M \rangle
\end{equation}

\section{The alignment cosine of the $^2\Sigma$ molecule in a magnetic field}
\label{app:cos2}

Within the Hund's~(b) basis functions, eq.~(\ref{WFfieldfree2}), the expectation value of the alignment cosine takes the form:
\begin{multline}
	\label{Cos2matrel}
	\langle \cos^2 \theta \rangle =
	a^2 (\omega_\text{m}) \bigl < N-\tfrac{1}{2} , \Omega, M \bigr | \cos^2 \theta \bigl | N-\tfrac{1}{2} , \Omega, M \bigr >  \\
	 +b^2 (\omega_\text{m}) \bigl < N+\tfrac{1}{2} , \Omega, M \bigr | \cos^2 \theta \bigl | N+\tfrac{1}{2} , \Omega, M \bigr > + 2 a (\omega_\text{m}) b (\omega_\text{m}) \bigl < N-\tfrac{1}{2} , \Omega, M \bigr | \cos^2 \theta \bigl | N+\tfrac{1}{2} , \Omega, M \bigr >,
\end{multline}
where the matrix elements of the $\cos^2 \theta$ operator in the Hund's case~(a) basis can be obtained from~(\ref{Phiz2JJ}) and~(\ref{Phiz2JJp1}):
\begin{equation}
	\label{Cos2JJ}
	\bigl < J, \Omega, M \bigr | \cos^2 \theta \bigl | J, \Omega, M \bigr > = \frac{1}{3} +
	\frac{2}{3}\frac{\left[ J(J+1) - 3M^2 \right] \left[ J(J+1) - 3\Omega^2 \right]}{J(J+1)(2J-1)(2J+3)}
\end{equation}
\begin{equation}
	\label{Cos2JJp1}
	\bigl < J, \Omega, M \bigr | \cos^2 \theta \bigl | J+1, \Omega, M \bigr > =2\Omega M \frac{\sqrt{\left[ (J+1)^2 - M^2 \right] \left[ (J+1)^2 - \Omega^2 \right]}}{J(J+1)(J+2)\sqrt{(2J+1)(2J+3)}}
\end{equation}
The coefficients $a (\omega_\text{m})$ and $b (\omega_\text{m})$ are given by the solution of the Zeeman problem, Eqs.~(\ref{HamMatr2x2})--(\ref{Psizeem}).

%========== R E F E R E N C E S =======================================

%============== T A B L E S =============================================

\newpage

\begin{table}[h]
\centering
\caption{The hybridization coefficients $a(\omega_\text{m})$ and $b(\omega_\text{m})$ for the $N=2,J=\frac{5}{2}, M$ state in the high-field limit, $\omega_\text{m} \gg \Delta E/B$, which arises for $\omega_\text{m} \gg 0.025$ for the $N=2$ level of the CaH$(X^2\Sigma^+)$ molecule. See text.}
\vspace{0.5cm}
\label{table:ab_coefs}
\begin{tabular}{| c | c | c |}
\hline 
\hline
$M$ &  $a(\omega_\text{m})$   & $b(\omega_\text{m})$  \\[3pt]
\hline  
$\frac{1}{2}$ &   $\sqrt{\frac{2}{5}}$  & $\sqrt{\frac{3}{5}}$  \\ [5pt]
-$\frac{1}{2}$ &  $\sqrt{\frac{3}{5}}$   & $\sqrt{\frac{2}{5}}$ \\[5pt]
$\frac{3}{2}$ &   $\frac{1}{\sqrt{5}}$    & $\frac{2}{\sqrt{5}}$ \\[5pt]
-$\frac{3}{2}$ &    $\frac{2}{\sqrt{5}}$   & $\frac{1}{\sqrt{5}}$ \\[5pt]
$\pm \frac{5}{2}$ &   0  & 1 \\[5pt]
 \hline
 \hline
\end{tabular}
\end{table}

\newpage

\begin{table}[h]
\centering
\caption{Hard-shell Legendre moments $\Xi_{\kappa 0}$ for He -- CaH ($X ^2 \Sigma^+$) and He -- O$_2$ ($X ^3 \Sigma^-$) potential energy surfaces at a collision energy of 200 cm$^{-1}$, and for He -- OH ($X ^2 \Pi_{\frac{3}{2}}$) potential at 1000 cm$^{-1}$.}
\vspace{0.5cm}
\label{table:legendre_coefs}
\begin{tabular}{| c | c | c | c|}
\hline 
\hline
& \multicolumn{3}{| c | }{$\Xi_{\kappa 0}$ (\AA) }   \\[3pt]
\hline
$\kappa$ &  He--CaH   & He--O$_2$  & He--OH \\[3pt]
\hline
 0 &   13.3207  & 9.5987 & 7.7941 \\
 1 &   -0.4397   & 0  & 0.1380 \\
 2 &   1.0140    &  0.5672 & 0.1625 \\
 3 &    0.6147   &0  & 0.0961 \\
 4 &   0.0337  & -0.1320 & 0.01789 \\
 5 &   -0.1475   & 0 & -0.0032 \\
 6 &     -0.0653   & 0.0250  & -0.0034 \\
 7 &     0.0265   & 0  & -0.0008 \\
 8 &     0.0277   & -0.0060  & 0.0002 \\
 \hline
 \hline
\end{tabular}
\end{table}

%============= F I G U R E S =========================================

\clearpage

\clearpage

\clearpage

\clearpage

\clearpage

\clearpage

\clearpage

\clearpage

\clearpage

\clearpage

\clearpage

\clearpage

\clearpage

\clearpage

\clearpage

\clearpage

\clearpage

\clearpage

\clearpage

\clearpage

\clearpage

\clearpage

\clearpage

\end{document}